\documentclass{article} 
\usepackage{graphicx}
\usepackage{amsmath}
\usepackage{amssymb}
\usepackage[ruled,linesnumbered]{algorithm2e}
\usepackage{mathtools}
\mathtoolsset{showonlyrefs}
\usepackage[caption=false]{subfig}
\usepackage{xcolor}

\newtheorem{theorem}{Theorem}

\newtheorem{lemma}{Lemma}

\newtheorem{remark}[theorem]{Remark}

\newtheorem{proposition}{Proposition}

\begin{document}

\title{Cooperative Predictive Cruise Control: A Bargaining Game Approach
}


\author{Miguel F. Arevalo-Castiblanco        \and
        Jaime Pachon \and
        Duvan Tellez-Castro \and
        Eduardo Mojica-Nava 
}



\date{Received: date / Accepted: date}

\maketitle

\begin{abstract}
This paper considers a cooperative cruise control problem from a predictive control perspective. Online decision-making is used to be executed during the driving process based on the information obtained from the network. We formalize a synchronization problem approach from a predictive control theory using bargaining games to find an operating agreement between the vehicles. Finally, we test these results in an emulation environment in a hardware-in-the-loop system.

\end{abstract}

\section{Introduction}
\label{intro}
In recent years, there has been an exponential increase in current vehicle transportation issues such as accidents, jam traffic, and air contamination, among others \cite{Jia2016,Van2007}. The main companies in the automotive industry have prioritized the development of vehicle control systems (e.g., 14\% of vehicles in the United States employed it recently). These control strategies mainly regulate the speed of a motor vehicle based on the presence of relatives in its vicinity \cite{Transport2014}. For this reason, using cooperative control strategies may reduce the transit time of vehicles on roads \cite{Mcdonald2016}.    

Recently, several cooperative control approaches have been proposed for networked vehicles \cite{Li2017}. The first works focused on cruise control without employing inter-vehicle communications to improve highway comfort and energy expenditure. The rise of wireless communications allows a network of vehicles to be seen as a cooperative study case \cite{Van2007}. The Cooperative Adaptive Cruise Control (CCAC) concept was developed using synchronization laws based on a reference acceleration profile \cite{Shladover2012}. The CCAC technique is based on string stability, where vehicles maintain a predefined distance at a constant speed, but it might not result in an optimal response \cite{Zohdy2016}. \\

On the other hand, predictive control strategies can find the optimal values in terms of energy or performance \cite{Yang2016}. In \cite{Farina2012,Trodden2017}, distributed predictive control strategies for agent synchronization have been developed using inter-agent communication and the construction of cost functions that involve network information. A Model Predictive Control (MPC) problem involving different agents where decisions depend on others can be considered as a game \cite{Grammatico2018}. Likewise, if these agents manage a common goal, the problem can be described as a bargaining game, where a set of players with a target modify their actions based on a disagreement between them \cite{Valencia2014}. Theoretically, the point of disagreement is defined as the minimum satisfaction expected for the negotiation. In practical implementation, cooperative cruise strategies can present several drawbacks, and only a few authors have validated these developments in highway vehicle platoons \cite{Chu2018}. The main practical tests of the theories explained are usually validated in dynamic emulation models, or robot networks \cite{Lin2018,Rayamajhi2018}.

The emulation of dynamic systems in hardware-in-the-loop enables the validation of real-time methods in situations where the models are unavailable (e.g., autonomous vehicle network). Emulation hardware based on embedded systems is used as a dynamic environment simulation with predefined models interaction and a visualization interface for validating desired signals \cite{Isermann1999}. For distributed networks, hardware emulation is commonly used in electrical and communication systems to validate future installations. The validation of these systems is mainly given by the need to prove the algorithm's correct operation in harsh application environments \cite{Zhang2016,Maniatopoulos2017}. In recent works, few results have been based on the theory of bargaining games as a problem of cooperative control in distributed systems, without considering vehicle platoon emulation \cite{Valencia2013}. Other works have emulated multi-agent systems without considering the application of predictive control techniques or game theory in vehicles \cite{nguyen2018}. 

Identifying the research gap falls at the application level in the case of cooperative cruise control, where the control strategy must solve a distributed optimization problem according to the states of the agents, the point of disagreement and the predefined cost functions for energy expenditure presets for simulation and emulation. A technique that allows managing control algorithms efficiently without increasing computational expense in the presence of the network is a distributed bargaining methodology. The main contribution of this paper then, is threefold: 1) First, the appropriation of a networked vehicle control problem is to be worked from a predictive perspective in a novel way. 2) Second, the inclusion of bargaining game theory for the cooperative control problem in simulation. This development allows it to be compared with centralized and decentralized predictive control algorithms for symmetric and non-symmetric cases. 3) Finally, a validation of the procedures developed in emulation with hardware-in-the-loop tools.

The rest of the paper is organized as follows: Section 2 presents the cooperative cruise control and bargaining games as a tool for Distributed Model Predictive Control (DMPC) background. Section 3 contextualizes the bargaining problem for the solution of the cooperative cruise problem. In Section 5, we show the study case application and its simulation. Section V shows the emulation of the system in hardware-in-the-loop, and finally, Section 6 sets out the conclusions of the work.
\section{Background}\label{S2}

This section presents the basic foundations of the cooperative cruise control theory, and DMPC as a bargaining game.

\subsection{Cooperative Cruise Control}
The cooperative cruise control problem has been extensively studied in latest years. Recent developments have focused on wireless vehicle-to-vehicle (V2V) communication that has grown commercially. The wireless communication led to the definition of the Grand Cooperative Driving Challenge (GCDC) to manage a platoon of vehicles that have this technology. The main objective of cooperative cruise control is synchronizing the vehicles on the road with the traffic profile considered by an established reference. The traffic profile is commonly constituted by the inter-vehicle distance and the speed on a highway, reducing vehicle time and fuel consumption. Communication is usually considered by the predecessor vehicle as a string stability case.

A simple description of the CACC setting (for longitudinal dynamics) is considered. In this case, each vehicle is modeled through its physical and mechanical parameters. The dynamics proposed in this case are linear as
\begin{align}
\Dot{p}(t)&=v(t), \\
\Dot{v}(t)&=a_{1}p(t)+a_{2}v(t)+b(\tau(t)+f(p,v)),
\label{dsystem}
\end{align}
where variables are speed $v$ and position $p$, respectively. The parameters $a_{1}$ and $a_{2}$ are transmission parameters, $b$ is related to transmission efficiency, and $\tau$ has the dimension of acceleration, or the force when it is multiplied by the vehicle mass. The term $f(p,v)$ is associated with an input uncertainty. These parameters can be included in the dynamics due to the approximation by an inversible steady-stable time-invariant model without the presence of uncertainties \cite{Filho2014}. This approach considerably reduces the complexity of the model without losing performance, and it has been used as an approximation in previous theory for validation of this type of problem \cite{baldi2019adaptive}.

\begin{figure}
    \centering
	\includegraphics[width=0.48 \textwidth]{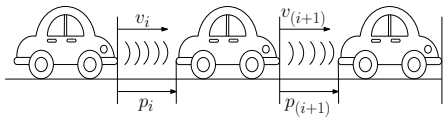}
	\caption{Vehicle Platoon Approach Diagram.}	
	\label{fig:carros}
\end{figure}

The controller must be able to regulate the speed of each vehicle and maintain a distance from its neighbors. The graphic representation of vehicles platoon is shown in Fig. \ref{fig:carros}, where the distance between each vehicle is defined as $d(t)=p_1(t)-p_2(t)$, that is the difference of the position of vehicles 1 and 2 with its respective subscripts $p_1$ and $p_2$.

Let the states of the system be $x(t)=[d(t),v(t)]^\top$, where the distance is position minus a predefined distance, and the control input $u(t)=\tau(t)$ with a possibly input matched uncertainty $f(p,v)$. It is possible to write the system as a linear affine continuous model of the form
\begin{equation}
\Dot{x}(t)=Ax(t)+b(u(t)+f(x)),
\label{Cdyn}
\end{equation}
with
\begin{equation}
A=\begin{bmatrix}
0 & 1\\
a_1 & a_2
\end{bmatrix},\quad b=
\begin{bmatrix}
0\\
b
\end{bmatrix}.
\label{matrix}
\end{equation}

This model considers the acceleration of neighboring vehicles in a $k-th$ instant of time. For implementation control prediction, it is necessary to use the discrete dynamics of the system. Therefore \eqref{Cdyn} is modified as
\begin{equation}
x(k+1)=A_kx(k)+b_k(u(k)+f(x_k)),
\label{LDM}
\end{equation}
with $A_k=e^{AT}$ and $b_k=A-1(A_k-I)b_k$.

Note that onboard sensors for controller action measure distance and speed, and the parameters of each vehicle can be different for a heterogeneous case.

\begin{remark}
    For the managed approximations of the cooperative cruise control theory, the considered model relates the position, inter-vehicular distance, and the speed of the vehicles in the network, assuming, in this case, a constant acceleration in an instant of time for each agent during development.
\end{remark}

\subsection{Distributed Model Predictive Control with Bargaining Games}

\begin{figure}[ht]
	\centering
	\includegraphics[width = 0.7\textwidth]{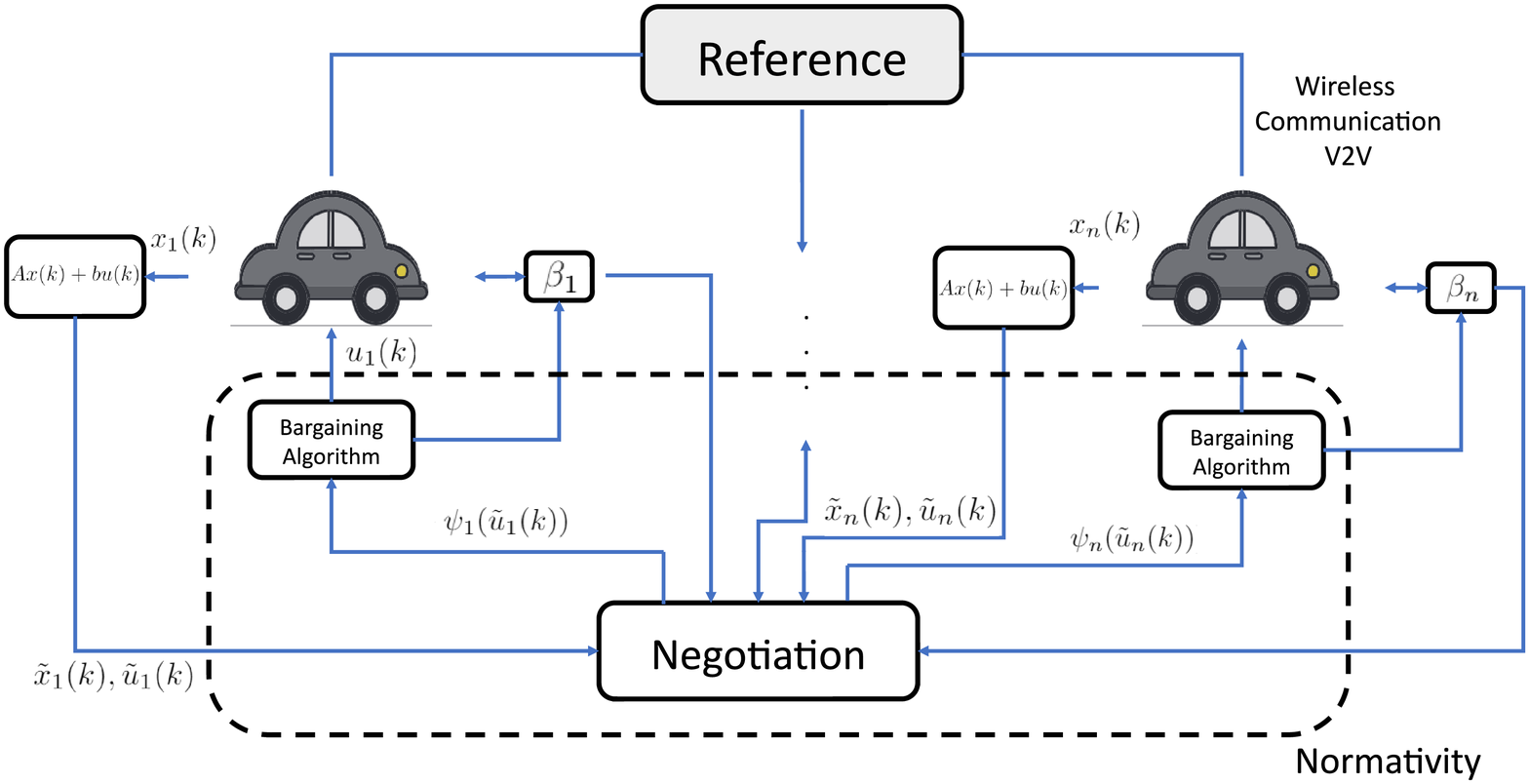}
	\caption{Model predictive control with bargaining games.}	
	\label{fig:blockBG}
\end{figure}

In this section, we introduce the basic concepts of MPC from a bargaining game perspective. It is contextualized as the negotiation method for solving a distributed optimization problem. The block diagram representing the framework is presented in Fig. \ref{fig:blockBG}, where it is observed that each vehicle, through its dynamics, enters a bargaining algorithm along a prediction horizon. This algorithm generates the optimal control action $u^*_i(k)$ together with the game relationship variable, the disagreement point $\beta_i$. This block diagram is used for non-symmetric bargaining cases. For symmetric cases, the characteristics of each vehicle are the same. All cases are regulated by regulatory aspects or physical restrictions of the vehicles to be considered.

A bargaining game is mathematically defined as the tuple $G=(N,\Lambda_i,\psi_i), \forall i \in N$, where $N$ is the number of vehicles involved in the process, $\Lambda$ is the decision space of the control law, and $\psi_i$ is defined as the local cost function of each vehicle. It is assumed that the vehicles are in a negotiating position to achieve a common objective, such as Nash's notions \cite{Nash1950}, where each player is sought to achieve the best benefit in common with their neighborhood according to the application \cite{Nash1953}. In the game, if it is impossible to reach an agreement, the term disagreement point is used for the bargain between vehicles \cite{Peters1992}.

Assuming the dynamics of each vehicle as in \eqref{Cdyn} and with its discrete representation \eqref{LDM}, the particular objective is to achieve energy-level optimization in each vehicle's operation. For this optimization problem, a usually quadratic cost function is defined in a distributed way as
\begin{equation}
L_i(\tilde{x}_i(k),\tilde{u}_i(k))=
\sum_{t=0}^{N_p}\psi_i(\tilde{x}_i(k),\tilde{u}_i(k)),
\label{scost}
\end{equation}
where $\tilde{x}_i(k)$ is the representation of vehicle $i$ states built along the prediction horizon $[x^\top_i(k),\ldots,x\top_i(k+N_p)]^\top$, likewise $\tilde{u}_i(k)=[u_i(k),\ldots,u_i(k+N_u),\ldots,u_i(k+N_p)]$ considering the control horizon $N_u$ with $N_u\leq N_p$. The local cost function is defined as
\begin{equation}
\psi_i(\tilde{u}(k))= \tilde{u}^\top(k)Q_{uui}\tilde{u}(k)+2x^\top_i(k)Q_{xui}\tilde{u}(k)+x_i^\top(k)Q_{xxi}x_i(k),
\label{CF_Local}
\end{equation}
which is positive defined, convex, and where $Q_{uui},Q_{xui}$ and $Q_{xxi}$ are weighted positive definite matrices, i.e., $Q_{uui}\succ 0$.

For the control problem formulation, it is important to define a decision space $\Lambda=\prod_{i=1}^N\Lambda_i$ for the whole system according to the physical operating conditions. An MPC problem with communication between agents is interpreted as a bargain so that it can be a bargaining game. For the analysis and solution of this type of game, Nash proposes an axiomatic methodology \cite{Nash1950b}, which was used in continuous and static systems \cite{Peters1992}.

A continuous representation for a bargaining game is with the tuple $(S,\beta_d)$, where $S$ is the game decision space, which is a non-empty closed subset of $\mathbb{R}^N$ and $\beta_d \in \text{int}(S)$ is the interaction disagreement point. For implementation purposes, it is important to consider the discrete dynamics of the game, so is then defined as ${(\Gamma(k),\beta_d(k))}^\infty_{k=0}$ with $\Gamma(k)$, a closed non-empty $\mathbb{R}^N$ subset that contains the profit functions values of each vehicle. The values of the states $x_i$, the set $\Gamma$, and the point of disagreement $\beta$ vary dynamically.

The evolution value of the disagreement point varies as
\begin{equation}
\beta_i(k+1) = \begin{cases}
\beta_i(k)-\mu(\beta_i(k)-\psi_i(\tilde{u}(k))) & \mbox{if } \beta_i(k)\geq\psi_i(\tilde{u}(k)), \\
\beta_i(k)+(\psi_i(\tilde{u}(k))-\psi_i(k)) & \mbox{if } \beta_i(k)<\psi_i(\tilde{u}(k)),
\end{cases}
\label{DPevolution}
\end{equation}
where $0\leq\mu\leq1$ is an adjustment constant according to the definition of the axioms of the negotiation processes raised from the work of John Nash \cite{peters1991characterizing}. In this case, if a vehicle decides to cooperate in the road, the disagreement point is reduced with a $\mu\left[\beta_i(k)-\psi_i(\tilde{u}(k))\right]$ factor, otherwise, it is increased by a  $\left[\psi(\tilde{u}(k))-\beta_i(k)\right]$ factor. As a complementary definition for the game, $\zeta_i$ is defined as the utopia point available for the vehicle $i$ as $\zeta_i(\Gamma)=\max{\psi_i}\text{ , }(\psi_i)_{i\in N}\in \Gamma$ , $\forall i \in N,$ $\Theta$ is defined as the union of the cost functions $\psi_i$ of the game, where then the discrete game can be interpreted as $\{\Theta,\beta(k)\}^\infty_{k=0}$.

\begin{remark}
	The analysis of a bargaining game can be carried out symmetrical for a game with similar characteristics between its players or non-symmetrical for a game where these characteristics differ, i.e., synchronization for oscillator systems with homogeneous characteristics or control of mechanical systems with heterogeneous physical characteristics.
\end{remark}

For the solution of a bargaining game, a non-symmetrical centralized scenario is proposed based on \cite{Borgers1997} as
\begin{align}
&\max_{\tilde{u}(k)}\sum_{i=1}^N(\lambda_i\log(\beta_i(k)-\psi_i(\tilde{u}_i(k)))), \\
\text{s.t. }&\beta_i(k)>\psi_i(\tilde{u}(k)),\\
&\tilde{u}(k)\in\Lambda,
\label{Doptimizacionns3}
\end{align} 
where $\lambda_i$ is a weight variable usually defined as $\lambda_i=\frac{1}{N}$, with $N$ as the number of vehicles involved in the process. However, for a distributed control analysis, the solution to the optimization problem is proposed as
\begin{align}
&\max_{\tilde{u}_i(k)}\sum_{r=1}^N(\log(\beta_r(k)-\kappa_r(\tilde{u}_i(k),\tilde{u}_{-i}(k)))), \\
\text{s.t. }&\beta_r(k)>\sigma_r(\tilde{u}_i(k),\tilde{u}_{-i}(k))\\
&\tilde{u}_i(k)\in\Lambda,
\label{Doptimizacionns4}
\end{align}
with $\kappa_r(\tilde{u}_i(k),\tilde{u}_{-i}(k))$ a distributed cost function usually defined quadratic and $\tilde{u}_{-i}(k)$ the set of the remaining vehicles control actions except for the agent $i$.

The optimization problem \eqref{Doptimizacionns4} differs from problem \eqref{Doptimizacionns3} about it considers $\tilde{u}_{-i}(k)$ fixed and only optimizes as a function of $\tilde{u}_i(k)$, this means that optimization does not involve the decisions of the entire network cooperatively. The solution to this problem then arises as a negotiation model that depends on the context given by the cooperative cruise control theory. This type of methodology does not use iterative solutions as others commonly used in distributed optimization problems \cite{Zou2019}, which reduces the computational cost in operation with great benefits in high-impact applications such as vehicle platoon. Bargaining methodology allows the solution to a distributed control problem by solving only one local optimization with the information collected by its neighbors and achieving an agreement based on the Nash theory of bargaining through the defined disagreement point. In summary, the objective is to apply a distributed control methodology for vehicles network on a highway. Based on the communication of their states, a negotiation can be interpreted as the solution to an optimization problem \eqref{Doptimizacionns4}.

\section{Cooperative Cruise Control as a Bargaining Game} \label{S3}

Considering the definition of a DMPC as a bargaining game presented in Section \ref{S2}, the cooperative cruise control problem is contextualized. In this scenario, the vehicles on a highway synchronize their dynamics from a reference model in vehicle distance and speed. The global cost function of the DMPC must be made up of two terms: one term associated with the tracking error in the distance between vehicles, and the other term with the speed of each one during its transit on the road.

The cooperative cruise control model is taken from \eqref{LDM}, where matrices $A_k, B_k$ are obtained from dynamic models and  \eqref{matrix}. The output is defined in this case as the speed of each vehicle. The vehicle's acceleration gives the control action $u(k)$. An operative constraint is defined according to comfort parameters of $32.6$ m/s \cite{Zoccali2018}.

Initially considering each vehicle independently, the local MPC problem is formulated as

\begin{align}
&\min_{\tilde{u}(k)}J\left(\tilde{u},x(k)\right) \\
\label{lop}
\text{s.t}.&x_i(k+N_p+1)=\bar{A}_ix_i(k+1)+\bar{b}_iu_i(k+1), \\
&y_i(k+N_p+1)=C_ix_i(k+1)+d_iu_i(k+1), \\
&x_i\in \mathcal{X}, \\
&u_i\in \Lambda,
\end{align}

with $N_u<\alpha<N_p-1$, $\Lambda$ is the decision space of the control law, $\bar{A}_i$ and $\bar{b}_i$ are the state matrix and vector resulting from the prediction along $N_p$. The characteristics of the software and hardware determine any possible delay that may be found in the communication for control operations.

Let $\kappa_i(\tilde{u}_i(k),\tilde{u}_{-i}(k))$ be the global cost function of each vehicle defined as
\begin{align}
\kappa_i(\tilde{u}_i(k),\tilde{u}_{-i}(k))=&\lambda|\tilde{v}_r(k)-\tilde{y}_v(\tilde{u}_i(k),\tilde{u}_{-i}(k))|\\
&+[\tilde{u}^\top_i(k),\tilde{u}^\top_{-i}(k)]\bar{H}_i[\tilde{u}^\top_i(k),\tilde{u}^\top_{-i}(k)]^\top+2\bar{F}[\tilde{u}^\top_i(k),\tilde{u}^\top_{-i}(k)]^\top,
\label{GCF}
\end{align}
where $H_i$ and $F_i$ are matrices obtained from $Q_{uu}$ and $Q_{ux}$, respectively. The input restrictions and states are time-independent and may differ for each vehicle. Therefore, the bargaining game for cooperative cruise control is defined as $G_{CCC}=\{N,\{\tilde{u}_i(k),\tilde{u}_{-i}(k)\},\Lambda_i\}$, $\forall i \in N$. Each vehicle at the control level has the same objective, to minimize the synchronization error to maintain the distance between vehicles and the speed in a stable state. The solution of this game with discrete characteristics of the form $\{\Theta,\beta(k)\}^\infty_{k=0}$ is solved by the following algorithm
\begin{algorithm}
	\KwResult{Optimal control signals}
	Initialize $u_{i}$,  $\beta_{i}$ \;
	\While{$e_{ij}>\Delta$}{
		$x_i$ send to others $x_i(k)$,$\beta_i(k)$\;
		$x_i$ solve \eqref{Doptimizacionns4}\;
		$x_i$ selects the first control action $\tilde{u}(k)$\;
		Each vehicle modifies $\beta_d$ according to \eqref{DPevolution}\;
		$x_i$ sends the modification of $\tilde{u}(k)$.
	}
	\caption{Distributed bargaining algorithm}
	\label{algorithm}
\end{algorithm}

Algorithm \ref{algorithm} methodologically explains that each vehicle sends its dynamics information to its neighbors as long as the synchronization error $e_{ij}$ is greater than a given constant $\Delta$. With this information, it is possible to solve the optimization problem \eqref{Doptimizacionns4} in each agent, to subsequently modify the values of the disagreement point until achieving convergence in the network synchronization. Finally, the modification to the point of disagreement is sent back to the neighbors. That process is one of the main contributions of this work since it summarizes the control methodology used in simulation and beyond with implementation for a cooperative cruise control problem.

The bargaining game result is mathematically defined as the tuple $\xi(\Theta,\beta(k))\\=\{\psi_1,\psi_2,\ldots,\psi_\mathcal{N}\}$ composed of the profit of each vehicle. If there is no cooperation on the highway, its value in the tuple is replaced by the disagreement point. 

\begin{proposition}
	The proposed solution $\xi(\Theta,\beta(k))$ of a discrete bargaining game ${(\Theta,\beta(k))}^\infty_{k=0}$ in $k$ steptime is unique and depends on optimization problem \eqref{lop} and $\Theta$ that must be convex.
\end{proposition}

\textit{Proof:} It follows by \cite{Valencia2013}.

As we mentioned in Remark 1, we consider two cases, one with similar characteristics of vehicles assimilating a symmetric game and another one with vehicles with non-similar characteristics assimilating a non-symmetrical game. However, we can note that non-symmetric bargaining games are the most commonly found in real-life applications since each subsystem that makes up a game usually has different characteristics than the others. The following lemma is proposed, based on Algorithm \ref{algorithm}.

\begin{lemma}
	Consider a cooperative cruise control problem as a bargaining game $\{(\Theta,\beta(k))\}_{k=0}^\infty$, then solution $\xi(\Theta,\beta(k))$ is the Nash bargaining solution at time $k$ computed by Algorithm \ref{algorithm}.
\end{lemma}

\textit{Proof:} According to the definition of the game and satisfying its axiomatic analysis [Section 2, \cite{Peters1992}], the solution of the cooperative cruise problem as a game $\{(\Theta,\beta(k))\}_{k=0}^\infty$ is defined as the Nash bargaining solution for every step time $k$ obtained through the negotiation problem $\blacksquare$.

Explicitly, this theory allows a complete structuring of the network, if required, through the transmission of the utility functions or system inputs to benefit the solution of each local optimization problem. These algorithms can make their own decisions separately, so their implementation does not need an iterative process. That decision-making process considerably reduces the computational burden that Lagrange multiplier-based solution methods can present.

\section{Simulation Results} \label{S4}
For the mentioned methodology, an application field is proposed based on the problem of an autonomous network of vehicles that is increasing nowadays, where each vehicle must follow the same patterns (position and speed). The most well-known technique for this problem is cooperative adaptive cruise control, an extension of the adaptive cruise control, a problem worked at a platoon level with onboard sensors. In this case, each agent is modeled as a linear second-order system such as in \eqref{LDM} with the remarks considered. For the experiment, the leading vehicle defines an acceleration profile that all agents must follow through a fixed distance between each one. This means in terms of synchronization $x_i-x_j\xrightarrow{}0$. It is important to highlight four aspects of this methodology, the vehicle dynamics, the distributed controller, the information communicated, and the graph communication topology \cite{Baldi2018Ifac}.

\begin{figure}[ht]
	\centering
	\includegraphics[width = 0.2\textwidth]{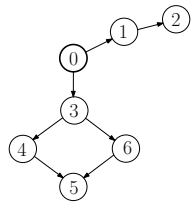}
	\caption{Study case communication graph.}	
	\label{fig:g1}
\end{figure}

A numerical simulation is performed to validate the proposed control laws. Figure \ref{fig:g1} shows the simulation digraph, where the agent $0$ acts as the leader node. The formation control idea in the platoon starts from the graph representing the communications, intercommunicated vehicles handle a speed and inter-vehicular distance contemplated by the reference vehicle.

%
\begin{table}
\centering
\caption{Players Coefficients and Initial Conditions.}
\label{tab:my_label}     
\begin{tabular}{lllll}
\hline\noalign{\smallskip}
 & $a_1$ & $a_2$ & $b_1$ & $x_0$  \\
\noalign{\smallskip}\hline\noalign{\smallskip}
$A_0$ & -0.25 & -0.5 & 1 & $[2 \hspace{0.1cm} 1]^\top$ \\  
			$A_1$ & -1.25 & 1 & 0.5 & $[1 \hspace{0.1cm} 2.1]^\top$ \\ 
			$A_2$ & -0.5 & 2.5 & 0.75 & $[1 \hspace{0.1cm} -0.2]^\top$ \\ 
			$A_3$ & -0.75 & 2 & 1.5 & $[3 \hspace{0.1cm}  2.3]^\top$ \\ 
			$A_4$ & -1.5 & 2.5 & 1 & $[3 \hspace{0.1cm}  0.6]^\top$ \\ 
			$A_5$ & -1 & 2 & 1 & $[2 \hspace{0.1cm}  -0.5]^\top$ \\ 
			$A_6$ & -0.75 & 1 & 0.5 & $[1 \hspace{0.1cm}  0.4]^\top$ \\	
\noalign{\smallskip}\hline
\end{tabular}
\end{table}

In Table \ref{tab:my_label} the simulation parameters used are presented, highlighting that these are used only for simulation, not for control design. The results of symmetric and non-symmetric games are shown in both simulation and implementation. MATLAB software and the \textit{fmincon} optimization problem-solving command are used for the optimization problems. The cost functions and decision variables are packed along the prediction horizon and using Kronecker-like structures for all the procedures. External parameters of communication or interaction between the agents are not considered for simulation purposes.

\subsection{Symmetric Game}

For the simulation of the system, symmetric and non-symmetric cases are presented. In the symmetric case, the cost function is defined as \eqref{GCF}, making a grouping according to the theory, and the local cost function is defined as \eqref{CF_Local}.

\begin{figure}[ht]
\centering
\begin{minipage}[b]{0.3\linewidth}
\includegraphics[width = \textwidth]{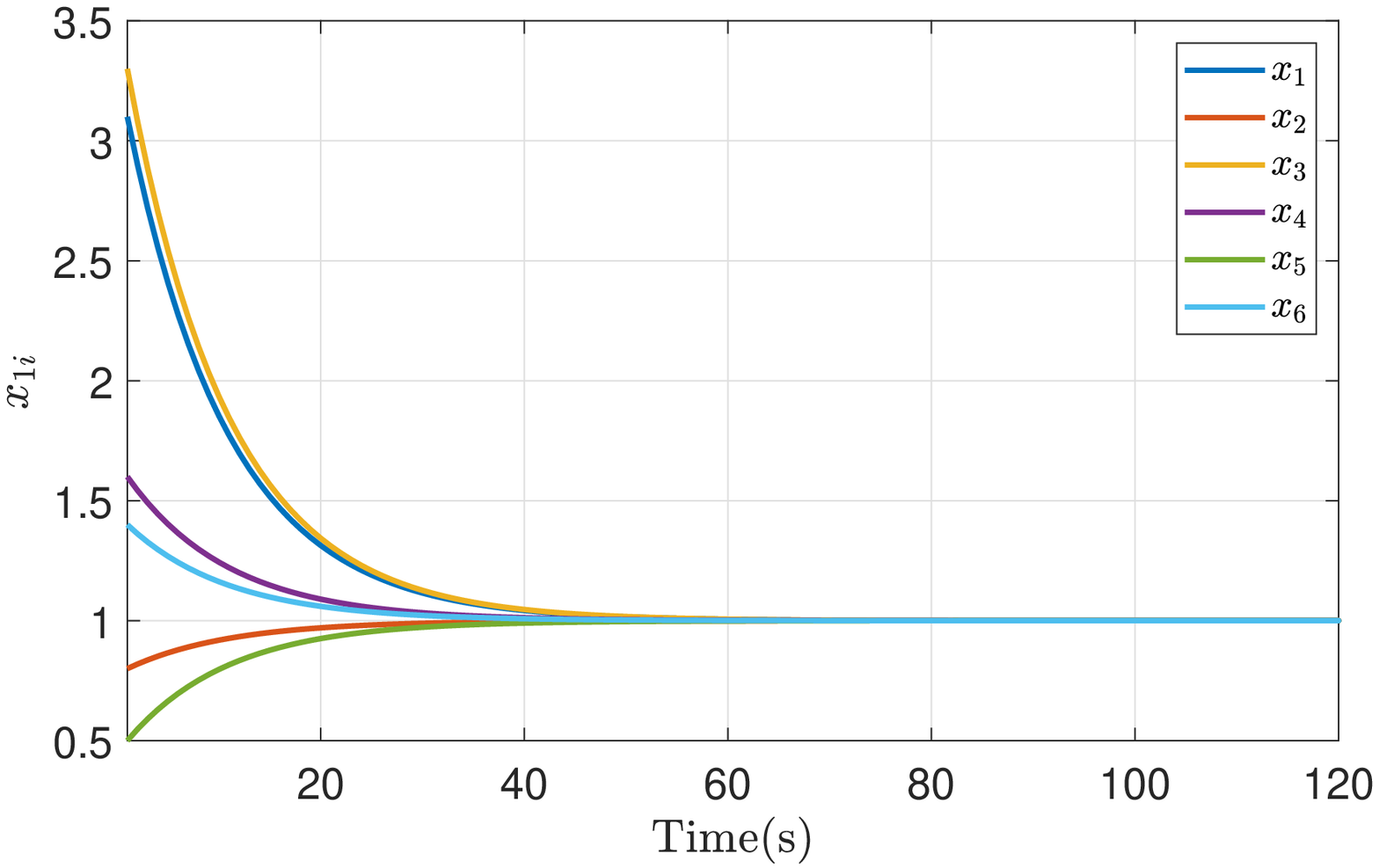}
\caption{Symmetric bargaining simulation standard game result}
\label{fig:comparacion1}
\end{minipage}
\quad
\begin{minipage}[b]{0.3\linewidth}
\includegraphics[width = \textwidth]{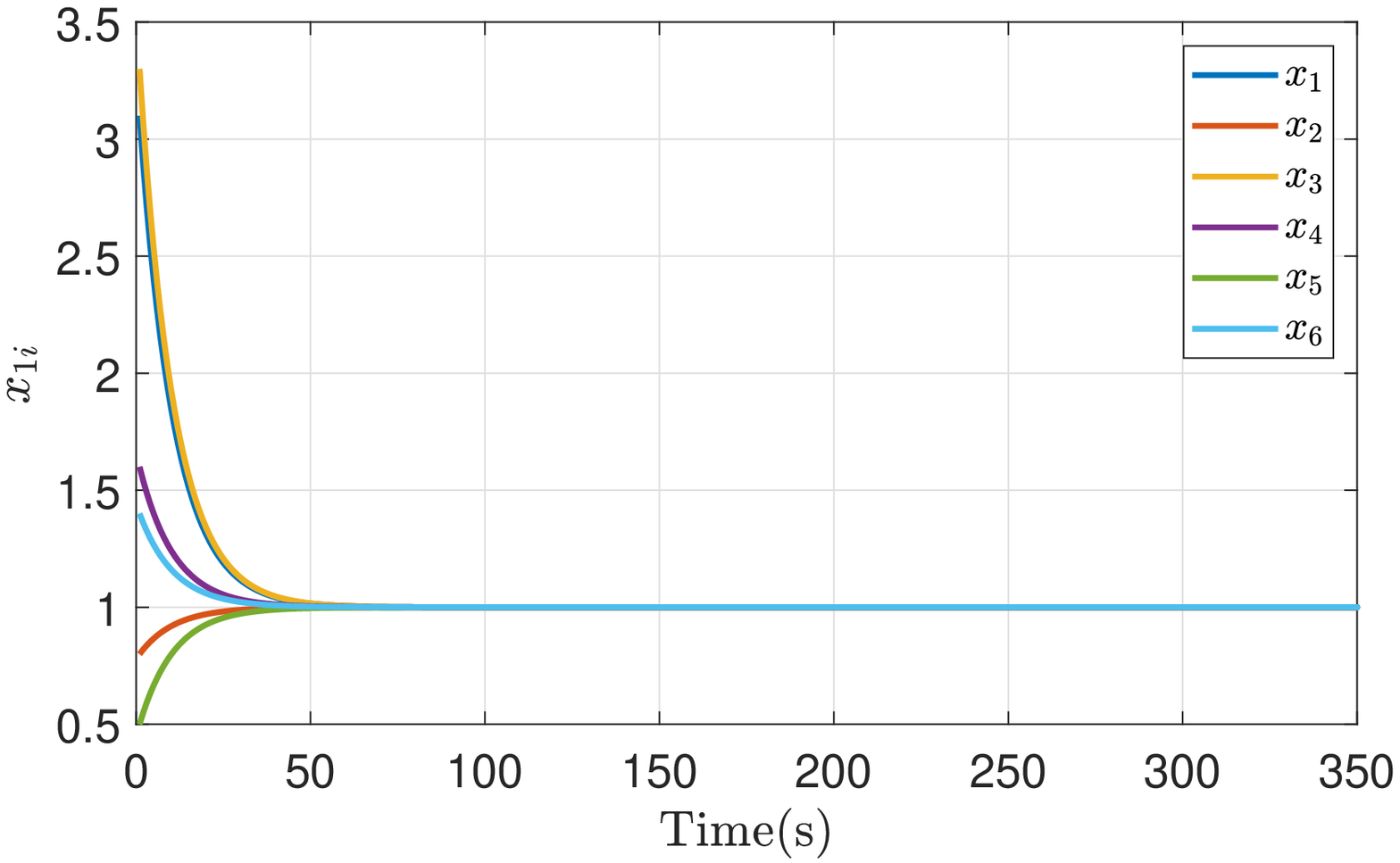}
\caption{Symmetric bargaining simulation centralized result}
\label{fig:comparacion1a}
\end{minipage}
\quad
\begin{minipage}[b]{0.3\linewidth}
\includegraphics[width = \textwidth]{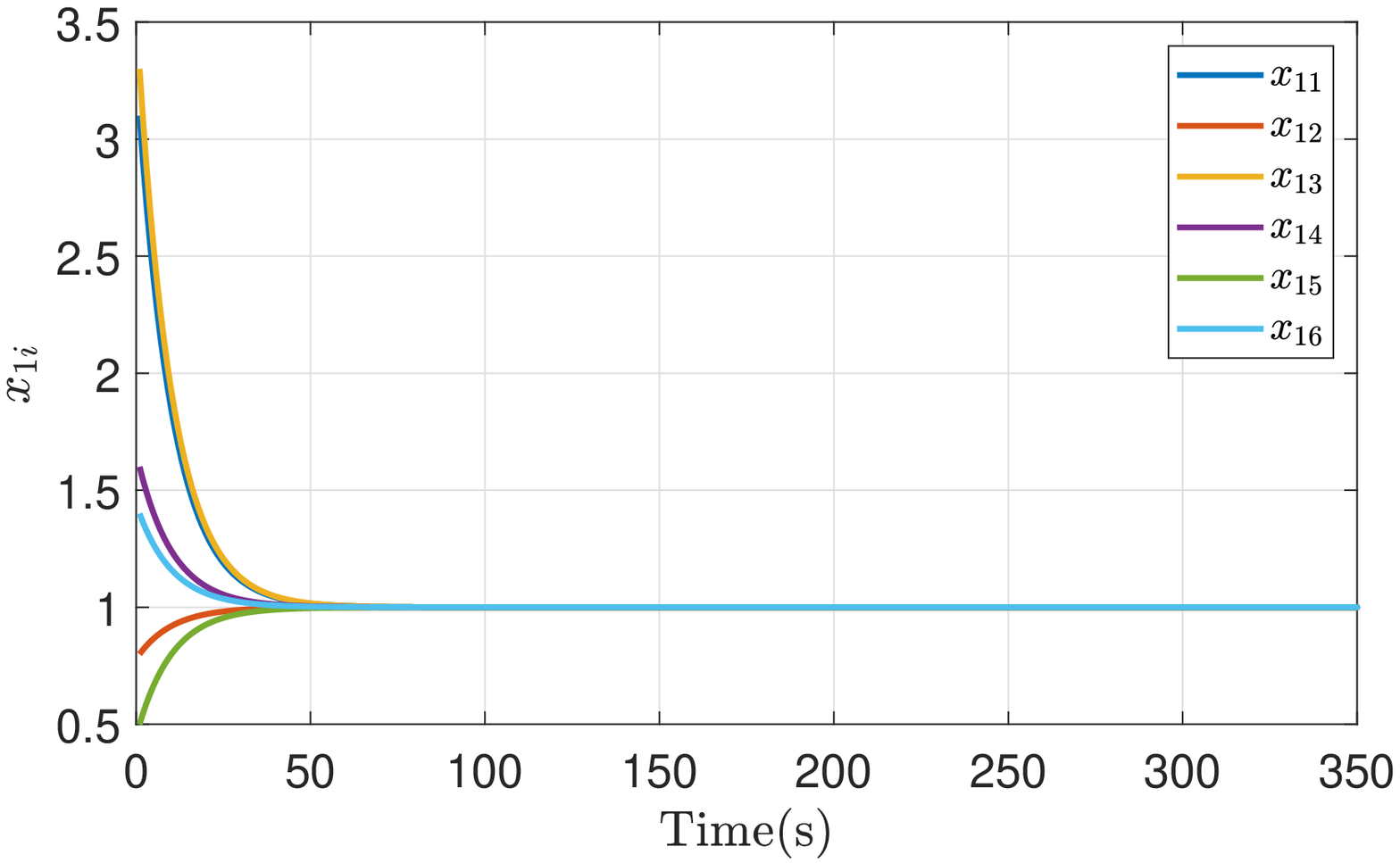}
\caption{Symmetric bargaining simulation decentralized result}
\label{fig:comparacion1b}
\end{minipage}
\end{figure}

The following figures show the agents' response when the solution to the distributed optimization problem is obtained. For simulation purposes, the agents dynamic is defined with the parameters $a_1=1$ and $a_2=b_1=-1$. Fig. \ref{fig:comparacion1} shows the network response with the bargaining model performed. 

\begin{figure}[ht]
\centering
\begin{minipage}[b]{0.5\linewidth}
\includegraphics[width = \textwidth]{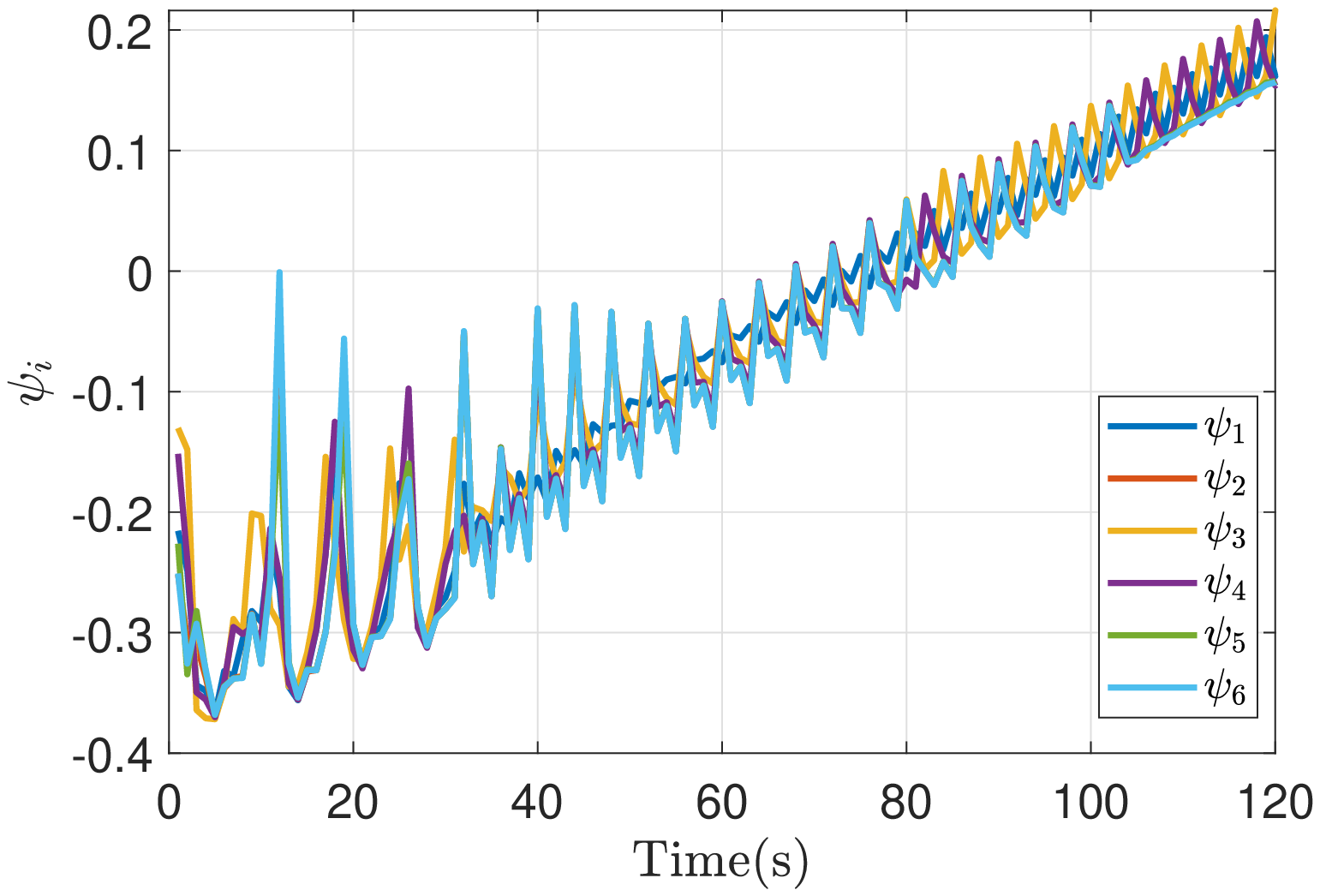}
\caption{Symmetric bargaining Cost Function}
\label{fig:4a}
\end{minipage}
\quad
\begin{minipage}[b]{0.5\linewidth}
\includegraphics[width = 1.2\textwidth]{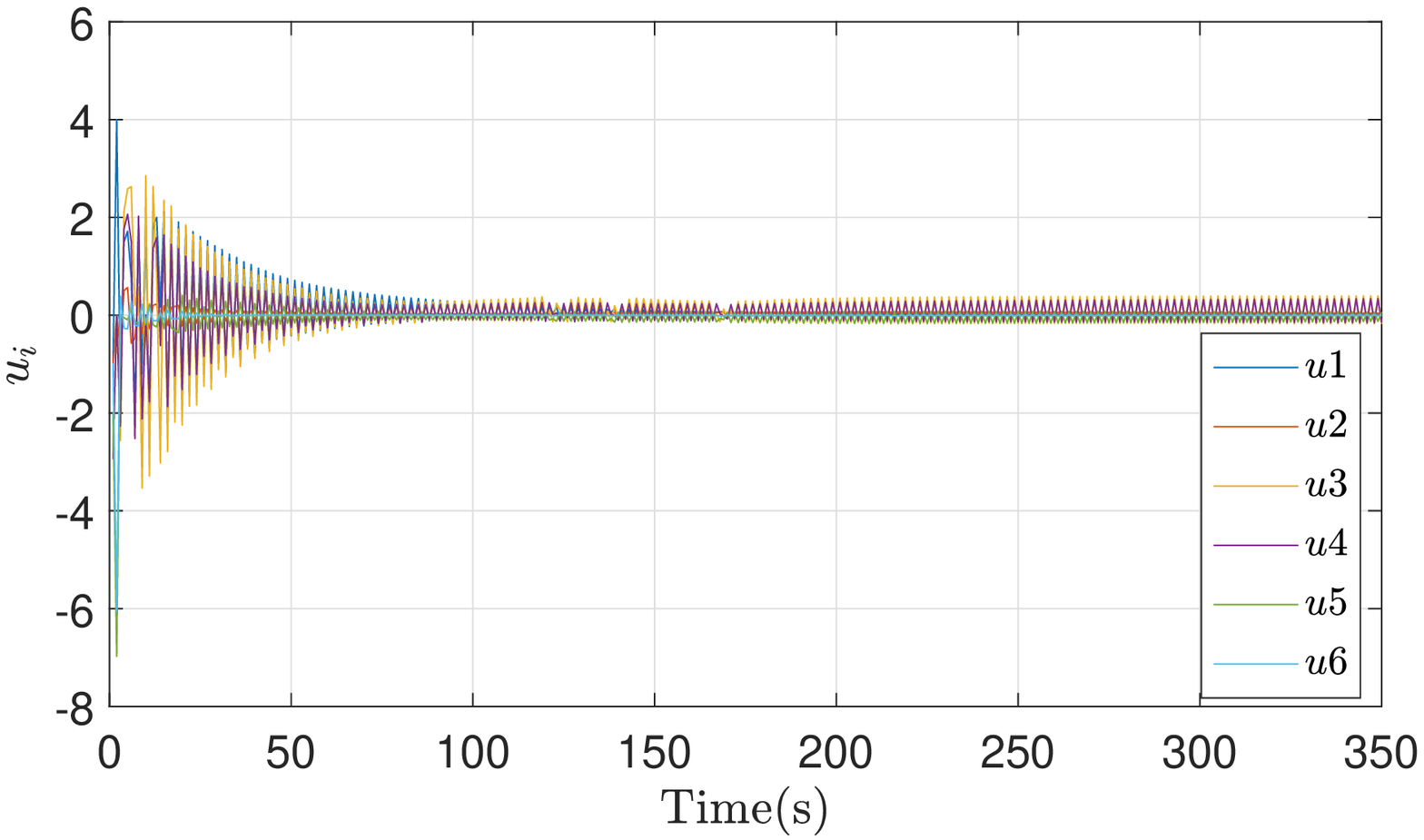}
\caption{Symmetric bargaining Control action}
\end{minipage}

\end{figure}

Similarly, Fig.\ref{fig:4a} shows the cost function and control action values evolution, respectively, where it is evident that they achieve synchronization based on Nash Theory.

We make a comparison with a centralized and decentralized model predictive control problem, reflected in Figure \ref{fig:comparacion1}--\ref{fig:comparacion1b}. Generally, a centralized problem is solved for each agent without any information sent. In a decentralized way, the information is sent, and with it, a single optimization problem per agent is solved. In the figures, it is evidenced that although the response of centralized and decentralized systems handles a faster response, synchronization is also achieved by communicating information in a distributed way, optimizing the processes of the applied network.


\subsection{Non-Symmetric Game}

\begin{figure}[ht]
\centering
\begin{minipage}[b]{0.3\linewidth}
\includegraphics[width = 0.9\textwidth]{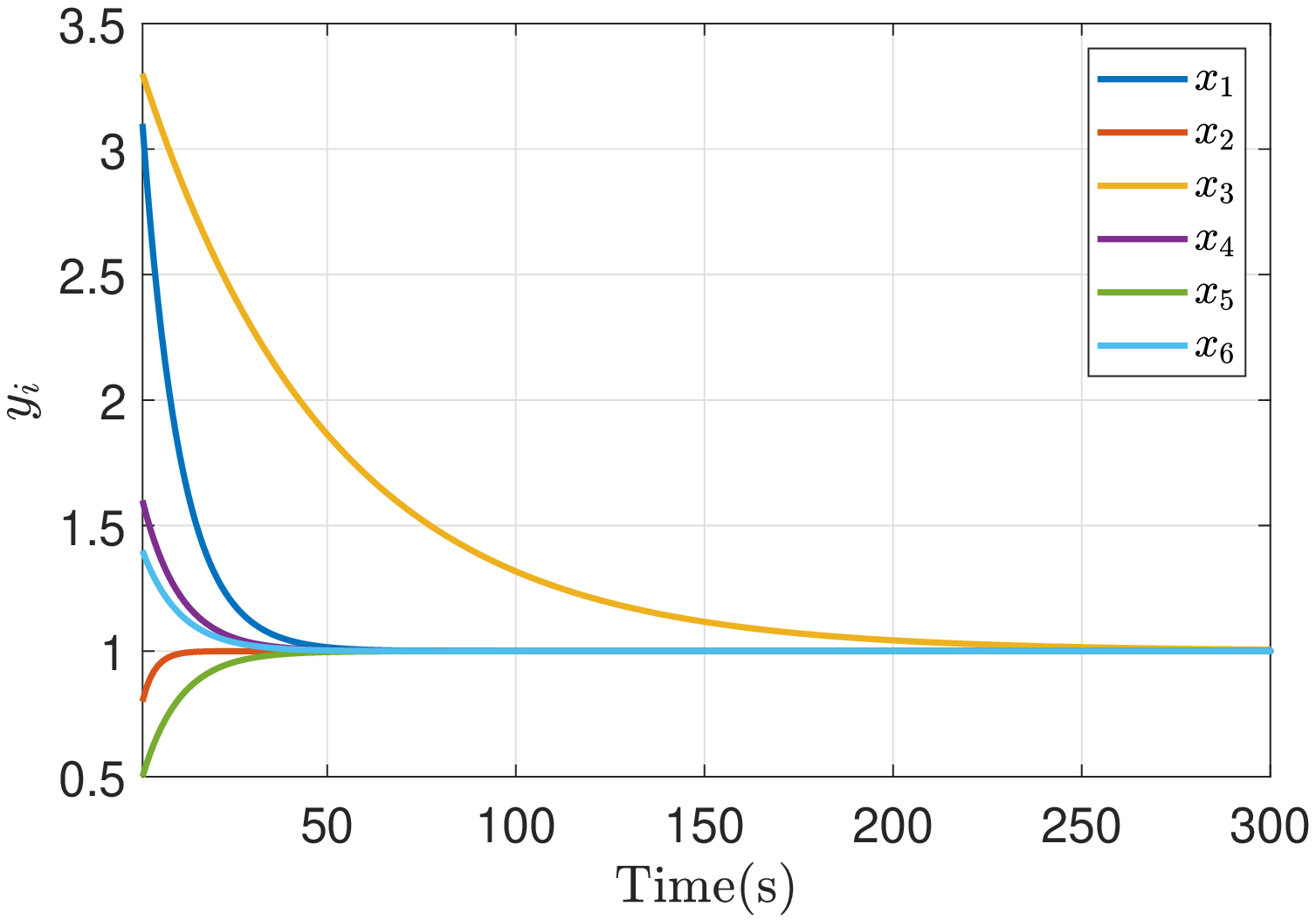}
\caption{Non-Symmetric bargaining simulation game result}
\label{fig:comparacion2}
\end{minipage}
\quad
\begin{minipage}[b]{0.3\linewidth}
\includegraphics[width = \textwidth]{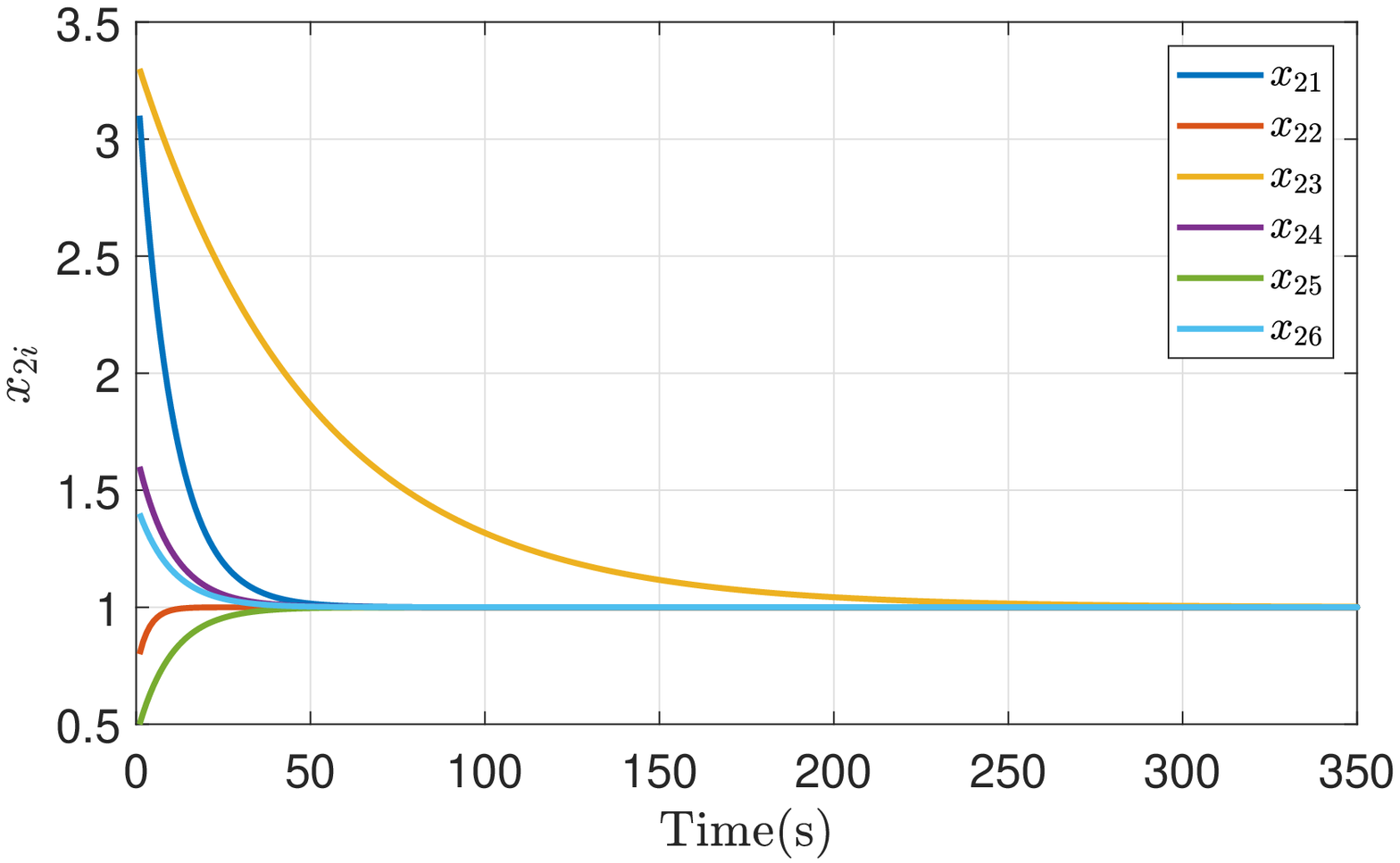}
\caption{Non-Symmetric bargaining simulation centralized result}
\label{fig:comparacion2a}
\end{minipage}
\quad
\begin{minipage}[b]{0.3\linewidth}
\includegraphics[width = \textwidth]{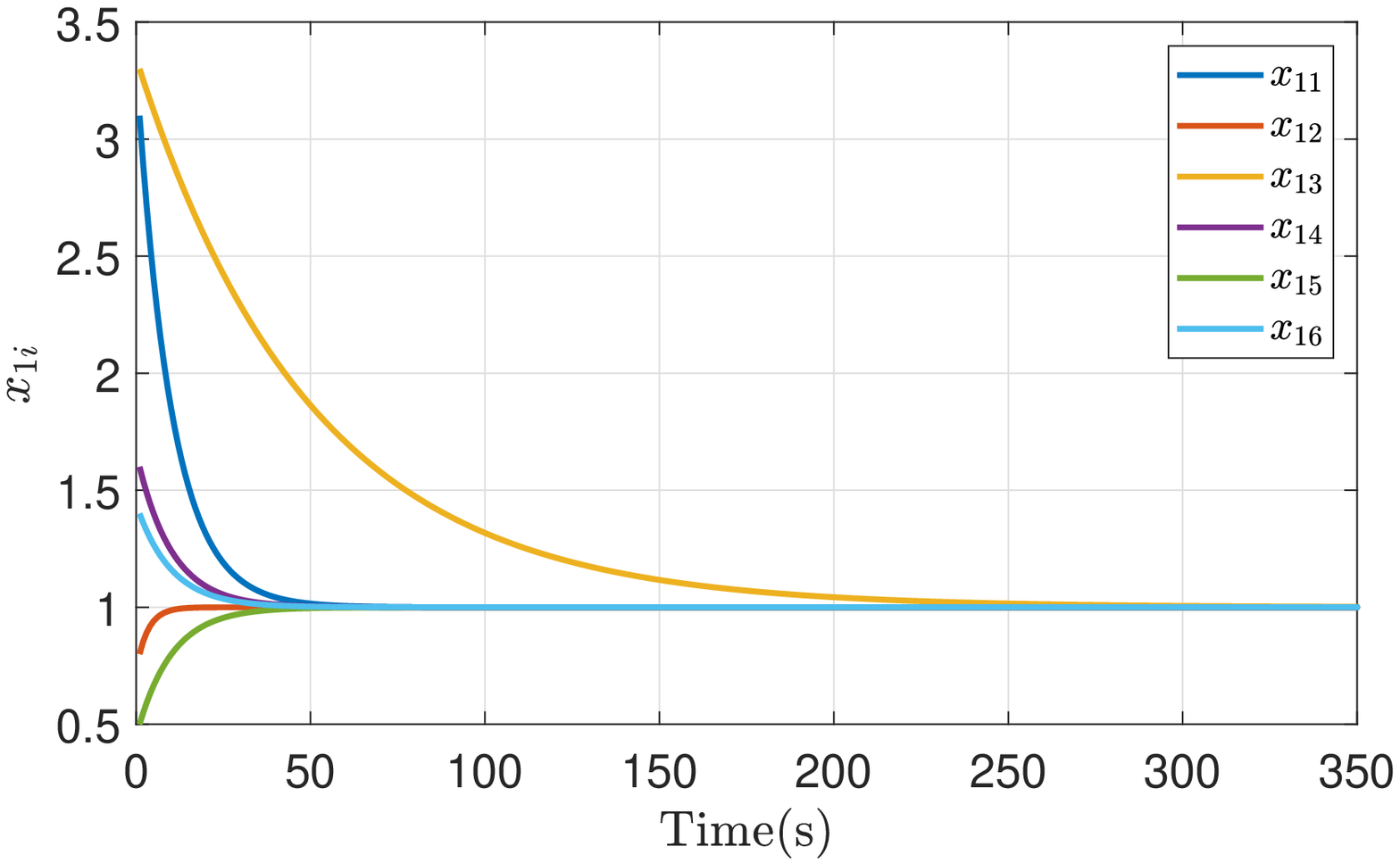}
\caption{Non-Symmetric bargaining simulation decentralized result}
\label{fig:comparacion2b}
\end{minipage}
\end{figure}

For the game with non-symmetric characteristics, the same case of cooperative cruise control is used. However, the parameters of each agent are taken from Table \ref{tab:my_label}. The cost function used is the same as in the symmetric case \eqref{CF_Local}. Fig. \ref{fig:comparacion2}--\ref{fig:comparacion2b} shows the output response of each system under the negotiation model compared with a centralized and decentralized predictive control methodology, where convergence is observed equally handling the information in a distributed way.

\begin{figure}[ht]
\centering
\begin{minipage}[b]{0.3\linewidth}
\includegraphics[width = 0.9\textwidth]{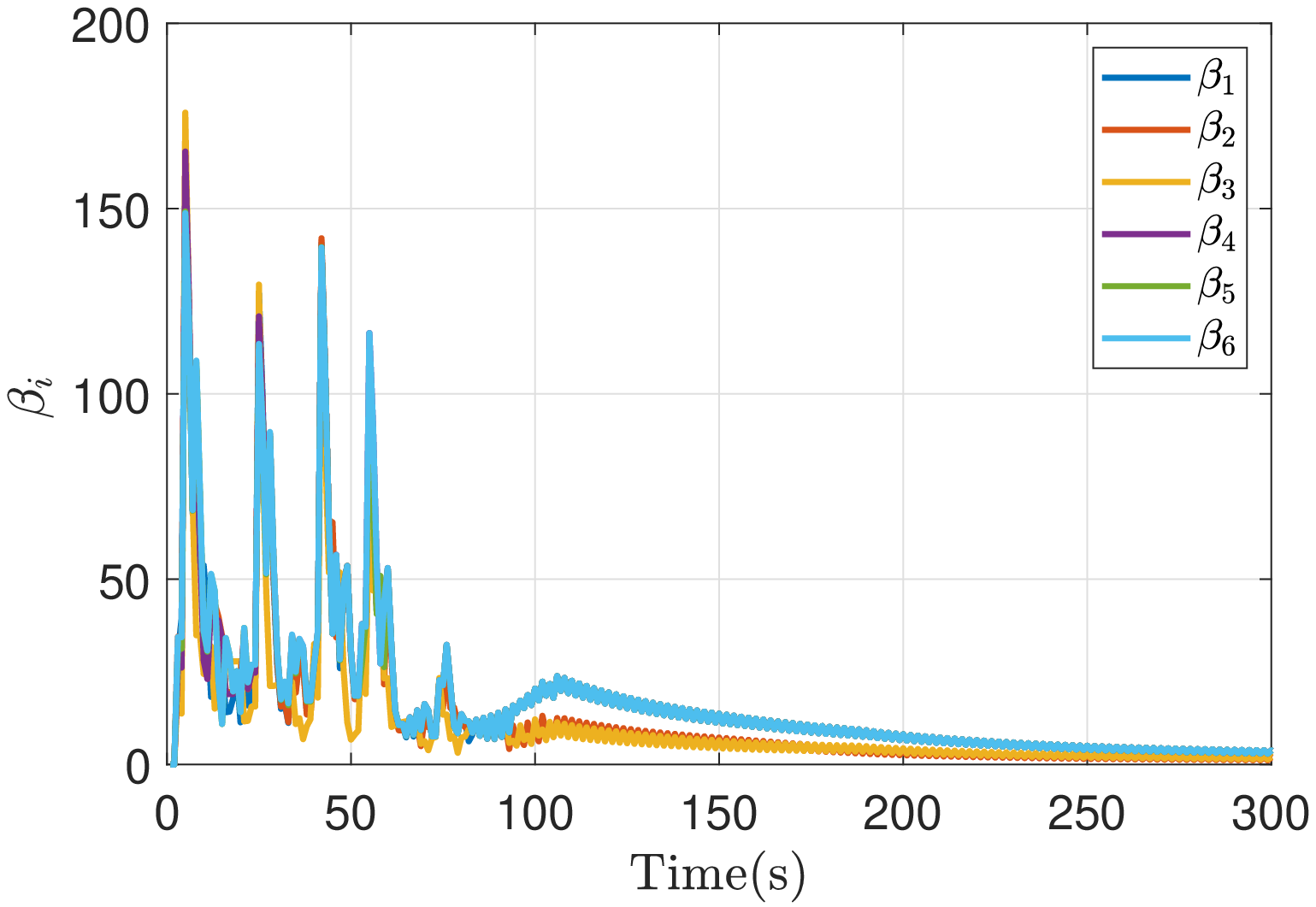}
\caption{Disagreement point Non-symmetric bargaining simulation result}
\label{fig:4-a}
\end{minipage}
\quad
\begin{minipage}[b]{0.3\linewidth}
\includegraphics[width = \textwidth]{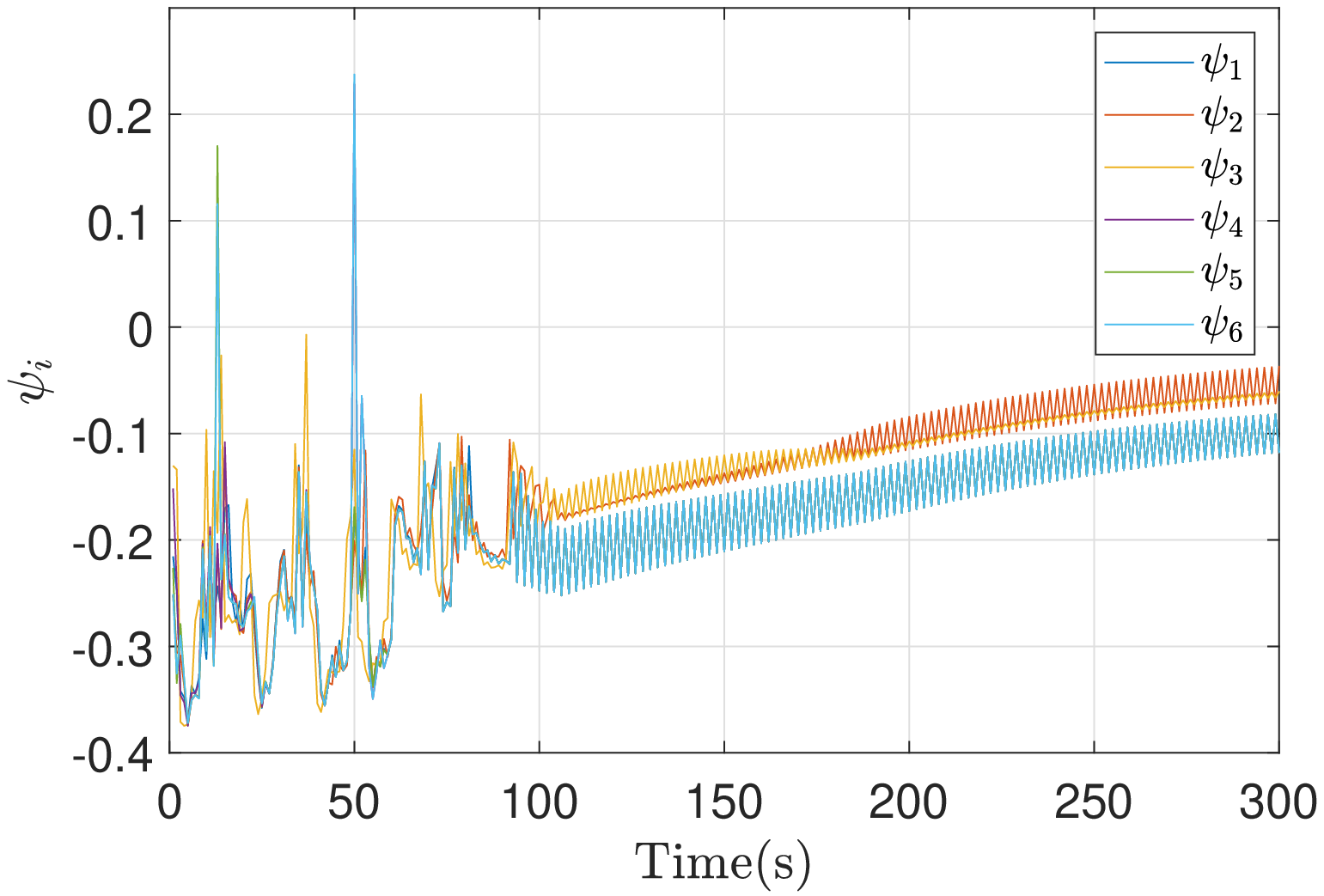}
\caption{Cost Function Non-symmetric bargaining simulation result.}
\label{fig:4-a1}
\end{minipage}
\quad
\begin{minipage}[b]{0.3\linewidth}
\includegraphics[width = \textwidth]{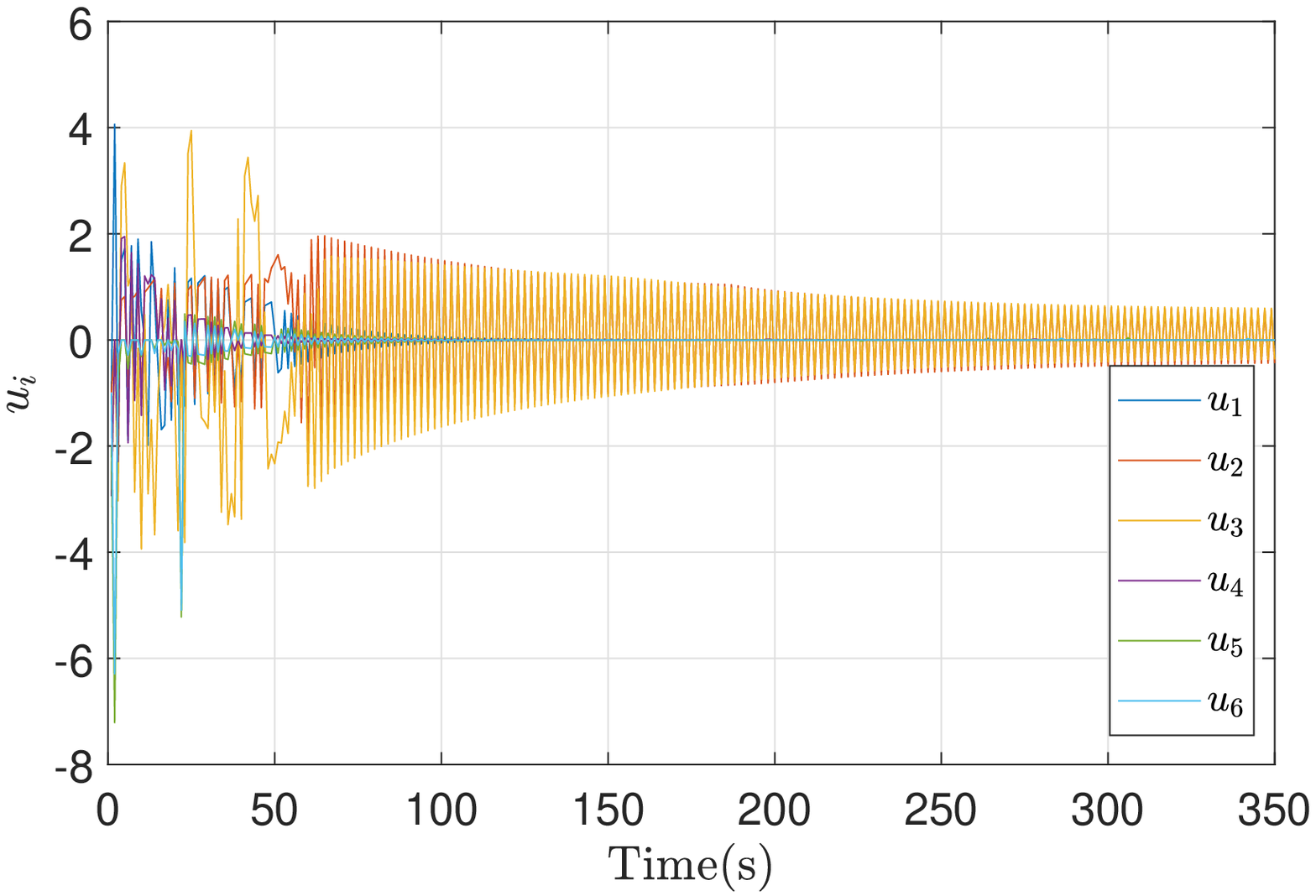}
\caption{Control action Non-symmetric bargaining simulation result.}
\label{fig:4-a2}
\end{minipage}
\end{figure}

In this case, the point of disagreement is plotted since, under the concept of symmetric and non-symmetric games, there may be a variation in said value (even considering symmetric cases with different initial conditions). Then, the response of the disagreement point is observed in Fig. \ref{fig:4-a} where the Nash agreement achieved from the consensus of this value is evidenced.

Finally, the cost function response is presented in Fig \ref{fig:4-a1}, where it is possible to validate the Nash theory through the convergence of these values in all agents, as well as the application of the control action for each one in the Fig. \ref{fig:4-a2}.

\section{Implementation Results} \label{S5}
\begin{figure}[ht]      
	\centering
	\includegraphics[width = 0.43\textwidth]{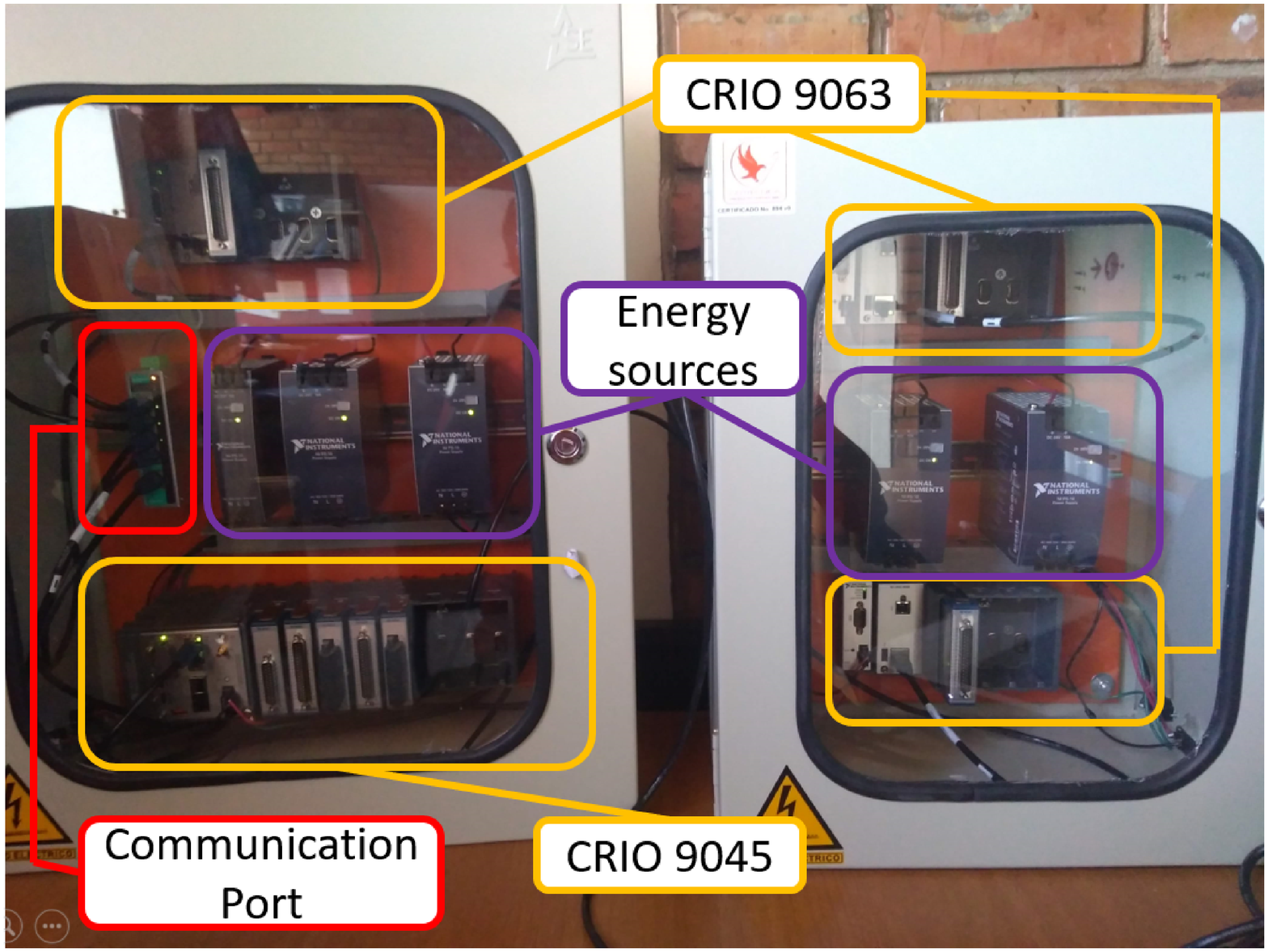}
	\caption{DESYNC lab implementation racks}
	\label{fig:foto}
\end{figure}

\begin{figure}[ht]      
	\centering
	\includegraphics[width = 0.65\textwidth]{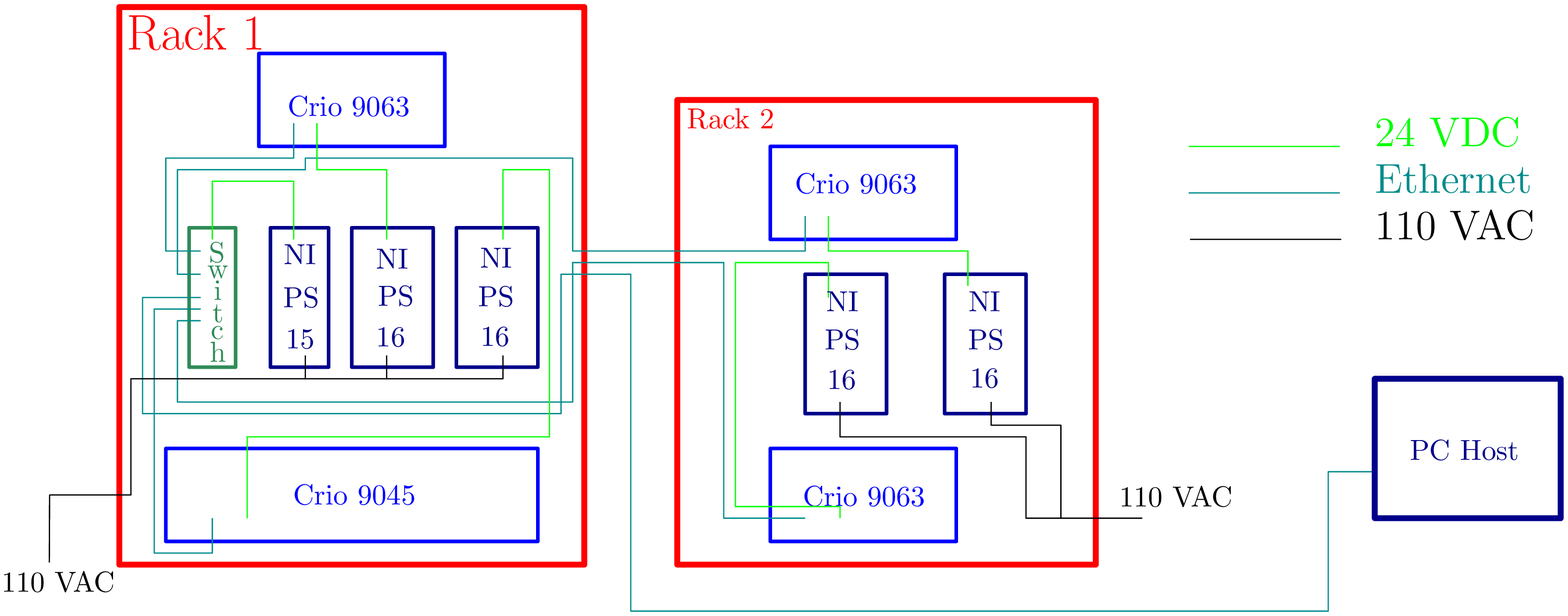}
	\caption{Block Rack Diagram for Implementation}
	\label{fig:Block}
\end{figure}

\begin{figure}[htbp]
	\centering
	\includegraphics[scale=0.5]{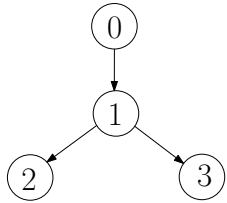}
	\caption{Communication graph used in emulation case}
	\label{ImplGraph}
\end{figure}

Basic experiments are performed to apply the developed algorithms with real-time simulation hardware. The validation is made through a temporal response of the developed algorithms, along with physical considerations. We use the National Instruments CompactRio controllers connected through an Ethernet network. Four controllers of two types are used for development, a NI9045 CompactRio controller and three NI9063 CompactRio controllers. The photo of the modules inside the DESYNC laboratory at Universidad Nacional de Colombia is shown in Fig. \ref{fig:foto}.
Similarly, as seen in Fig. \ref{fig:Block}, the inclusion of each controller within Rack is observed. Two NI9063 controllers and their two power supplies are observed. In the other cabinet, the remaining NI9063 controller and the NI9045 controller are shown next to their power supplies and the Ethernet communication switch. The communications graph is defined as in Fig. \ref{ImplGraph}.

Similarly, for the dynamic models implemented, the same dynamic \eqref{dsystem} is used with a \textit{Tustin} discretization for system matrices with a sample time of 0.1s. The simulation parameters are observed in Table \ref{tab:my_label2}.

\begin{table}
\centering
\caption{Agent's Coefficients and Initial Conditions for emulation case.}
\label{tab:my_label2}     
\begin{tabular}{ccccc}
\hline\noalign{\smallskip}
 & $a_1$ & $a_2$ & $b_1$ & $x_0$  \\
\noalign{\smallskip}\hline\noalign{\smallskip}
$A_0$ & -0.25 & -0.5 & 1 & $[1 \hspace{0.1cm} 2]^\top$ \\  
			$A_1$ & -1.25 & 1 & 0.5 & $[1 \hspace{0.1cm} 4]^\top$ \\ 
			$A_2$ & -0.5 & 2.5 & 0.75 & $[-1 \hspace{0.1cm} 2]^\top$ \\ 
			$A_3$ & -0.75 & 2 & 1.5 & $[1 \hspace{0.1cm}  4]^\top$ \\
\noalign{\smallskip}\hline
\end{tabular}
\end{table}

\begin{figure}[!htb]
	\centering
	\includegraphics[width = 0.5 \textwidth]{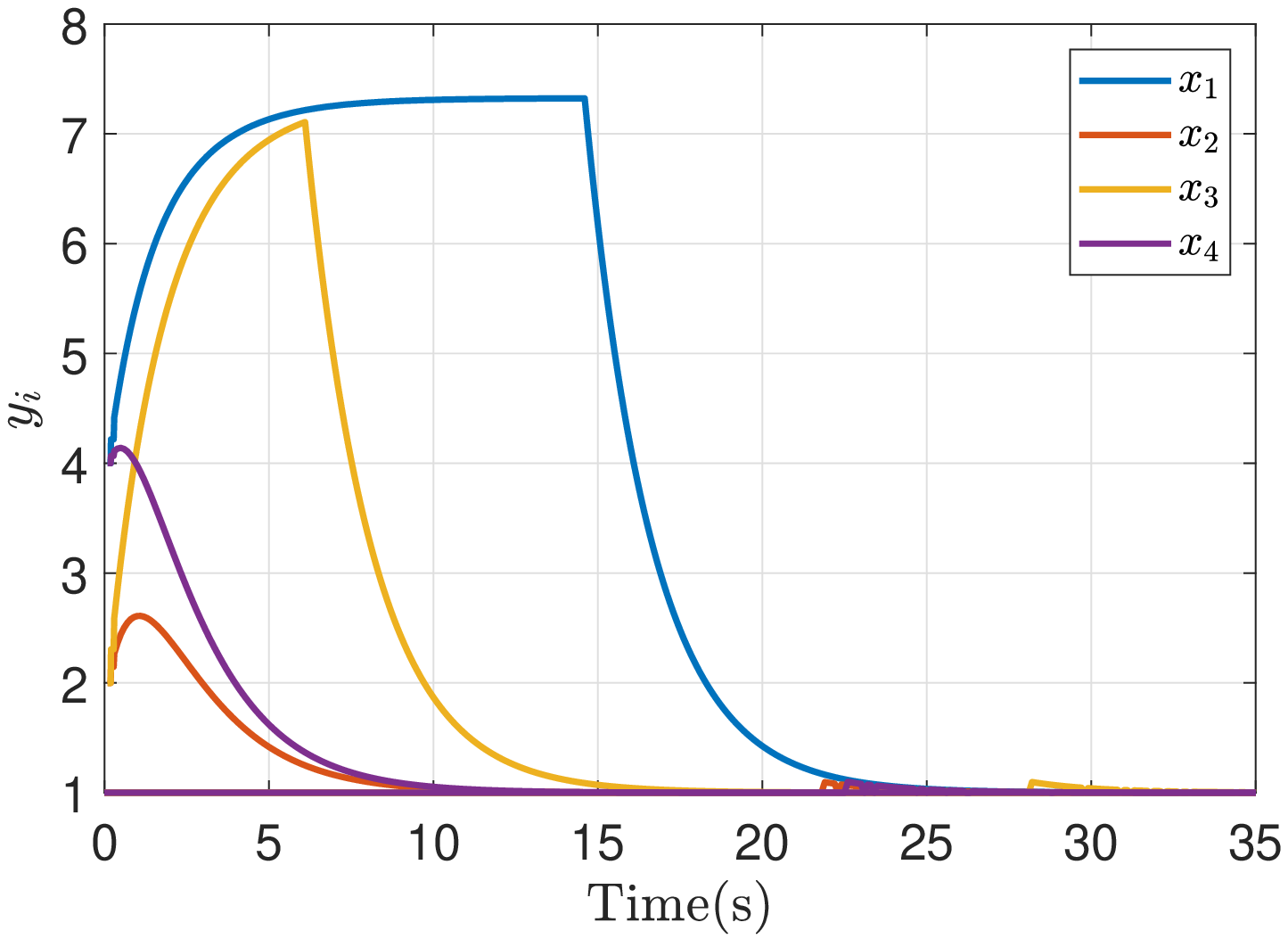}
	\caption{Output synchronization of bargaining game theory implementation in the symmetric case.}
	\label{fig:5-1}
\end{figure}

For the implementation, communication is made between Labview and Simulink, where the dynamics of the controllers are emulated. For the response of the symmetric game, the same parameters of the simulation case are used ($a_1=1$ and $a_2=b_1=-1$). Fig. \ref{fig:5-1} shows the controller's response implemented in a symmetric game. In emulation, the system presents some fluctuations initially, but their response also achieves an adequate bargain. In all those cases, the fluctuations are derived from the fact that by embedding the dynamics and control in each module separately and requiring communication between them, the response does not handle the same synchronization interval as in simulation, where the communication does not have delays.

\begin{figure}[ht]
\centering
\begin{minipage}[b]{0.5\linewidth}
\includegraphics[width = \textwidth]{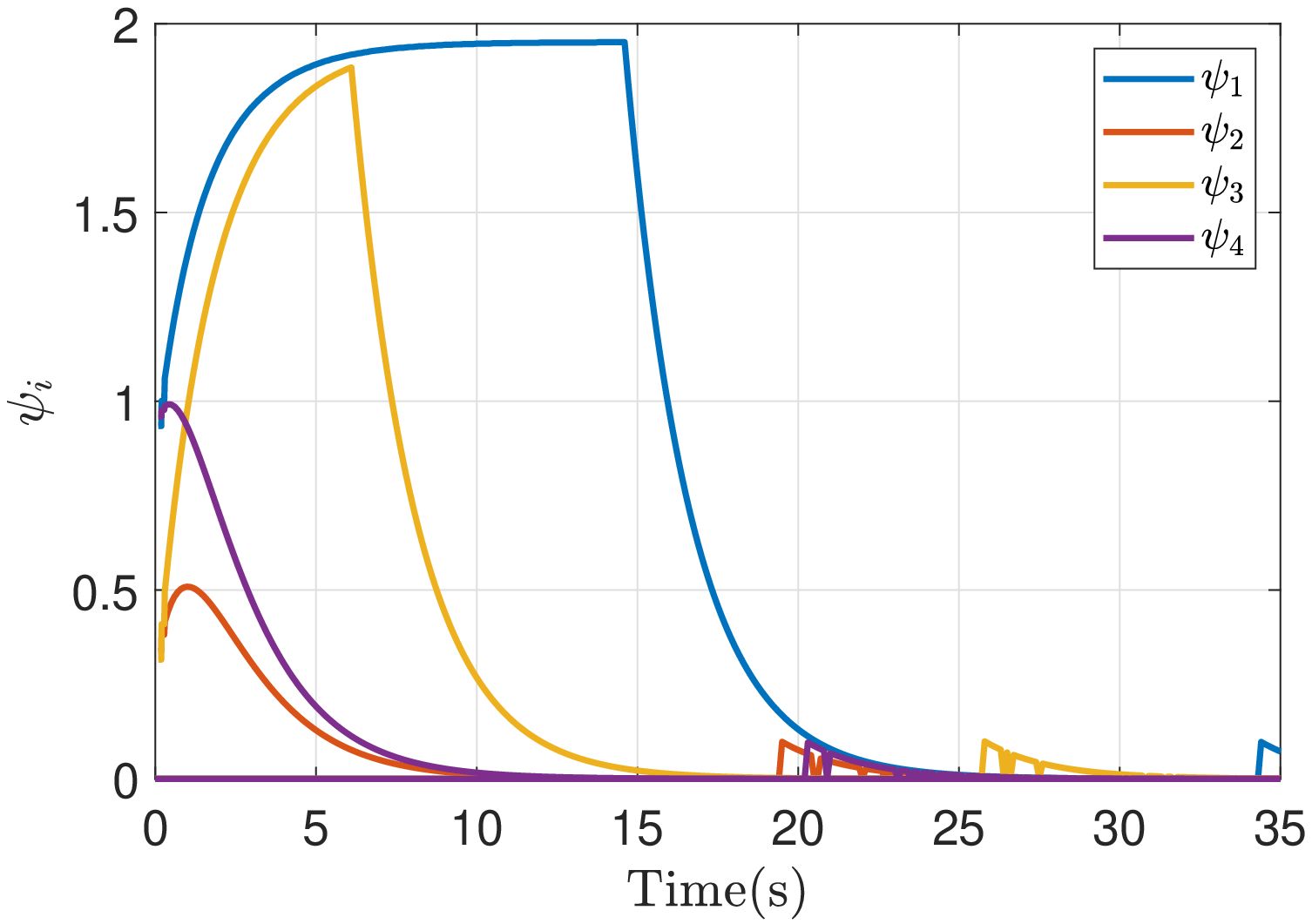}
\caption{Cost function of bargaining game theory implementation in a symmetric case.}
\label{fig:5-2a}
\end{minipage}
\begin{minipage}[b]{0.5\linewidth}
\includegraphics[width = \textwidth]{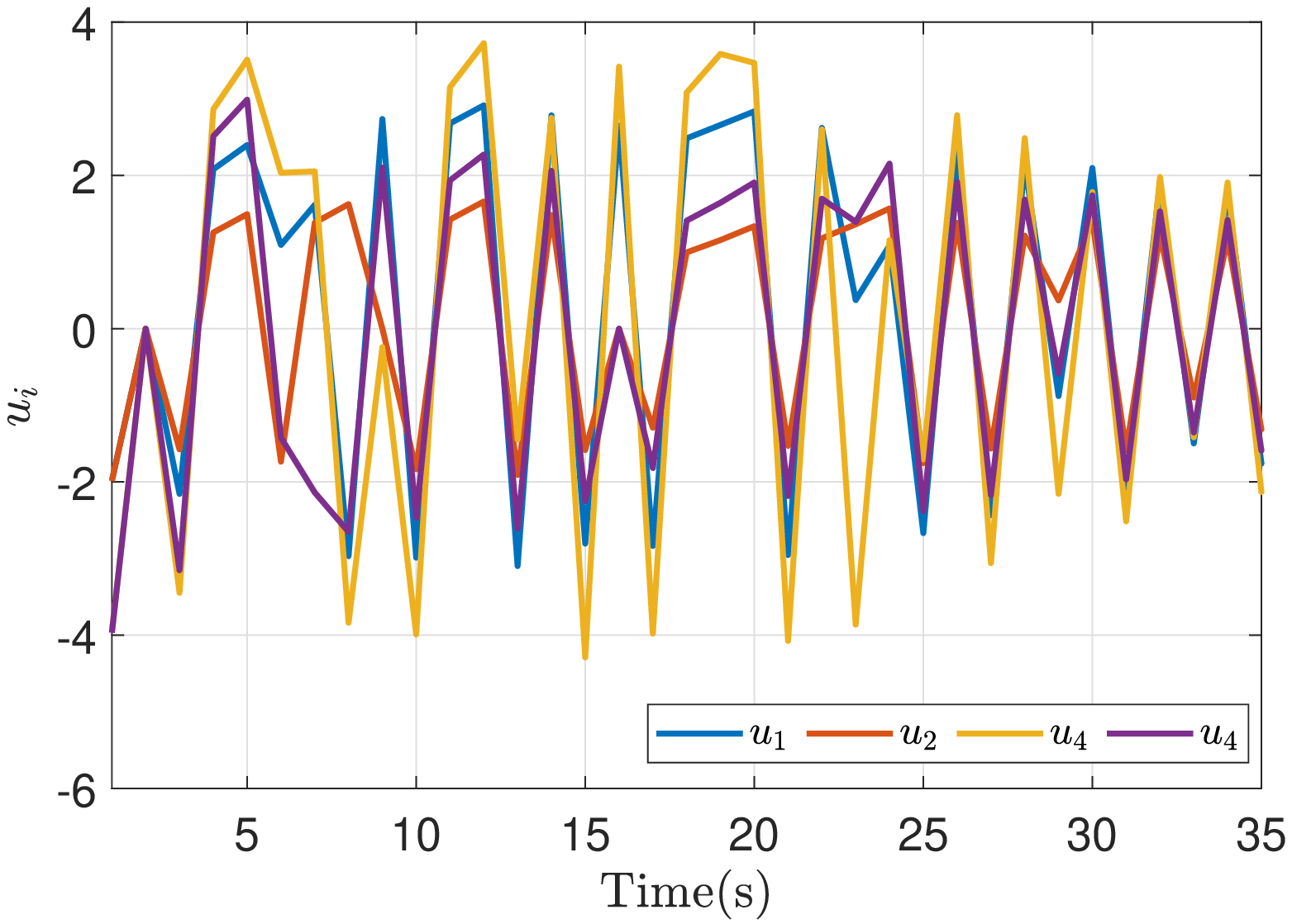}
\caption{Control action of bargaining game theory implementation in a symmetric case.}
\label{fig:5-2b}
\end{minipage}
\end{figure}

Similarly, the response of the agents' cost function is observed in Fig. \ref{fig:5-2a}, which maintains similarity with the response of the simulation case and also reaches a correct bargain, as well as the application of the control action in the Fig. \ref{fig:5-2b}.

\begin{figure}[!htb]
	\centering
	\includegraphics[width = 0.47 \textwidth]{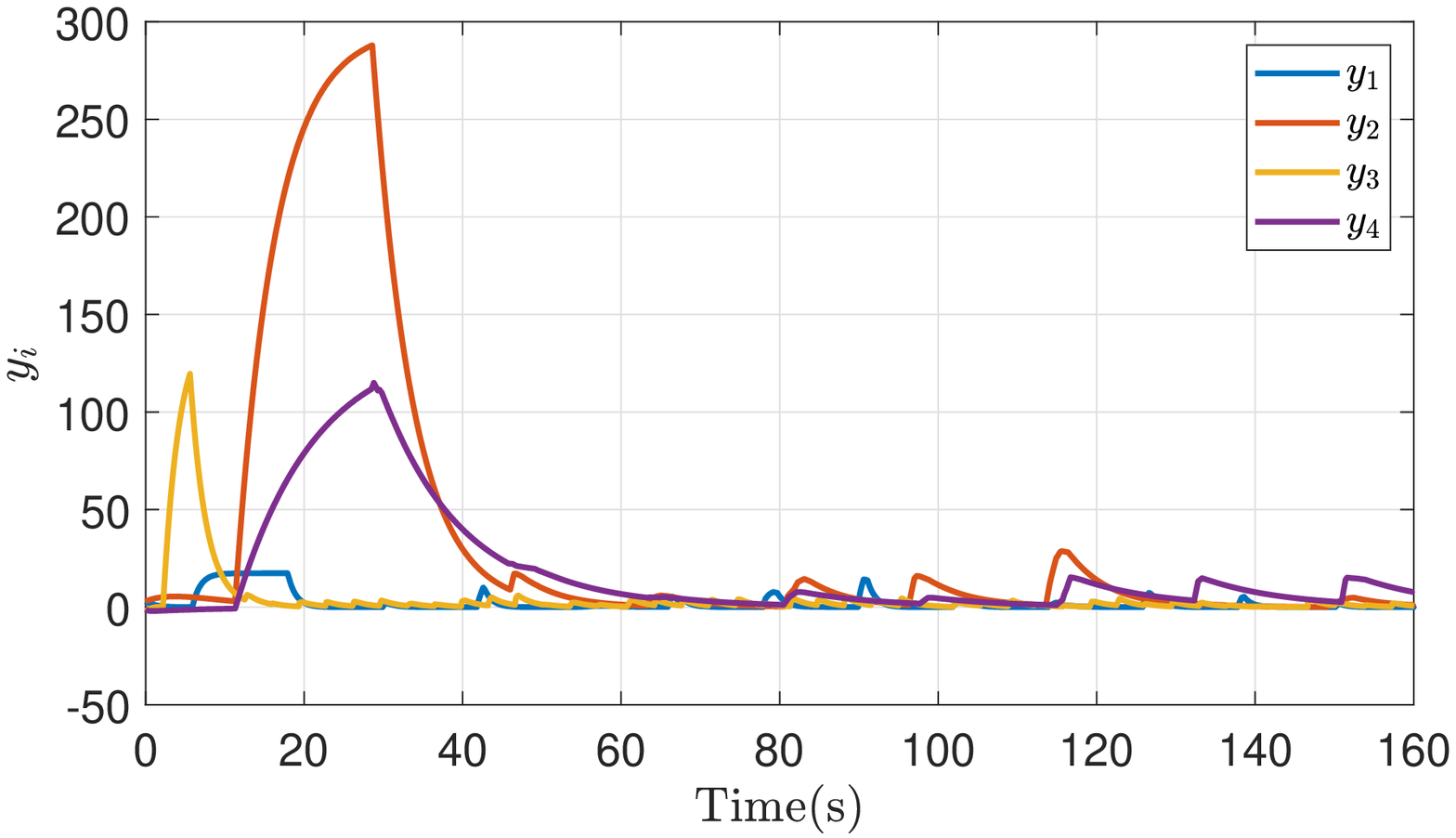}
	\caption{Output synchronization of bargaining game theory implementation in a  Non-symmetric case.}	
	\label{fig:5-4}
\end{figure}

The response of the controller to non-symmetric cases is also validated, according to the parameters of Table \ref{tab:my_label2}, Fig. \ref{fig:5-4} shows the response of the agents' output in this case, where it is evidenced that as in the previous cases, the system achieves correct bargaining in a distributed scenario.

\begin{figure}[ht]
\centering
\begin{minipage}[b]{0.3\linewidth}
\includegraphics[width = \textwidth]{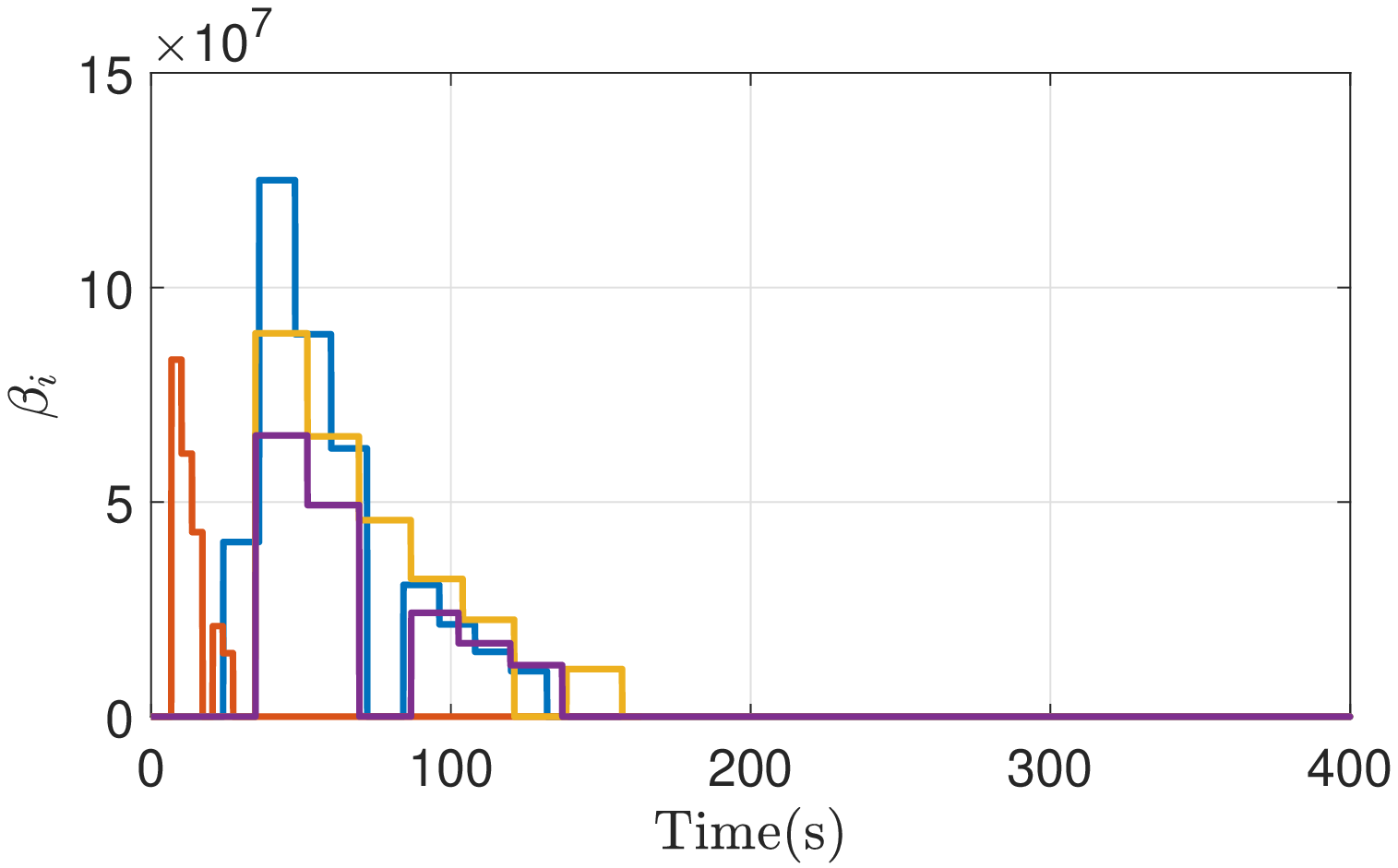}
\caption{Disagreement point of bargaining game theory implementation in a non-symmetric case.}
\label{fig:5-5}
\end{minipage}
\quad
\begin{minipage}[b]{0.3\linewidth}
\includegraphics[width = \textwidth]{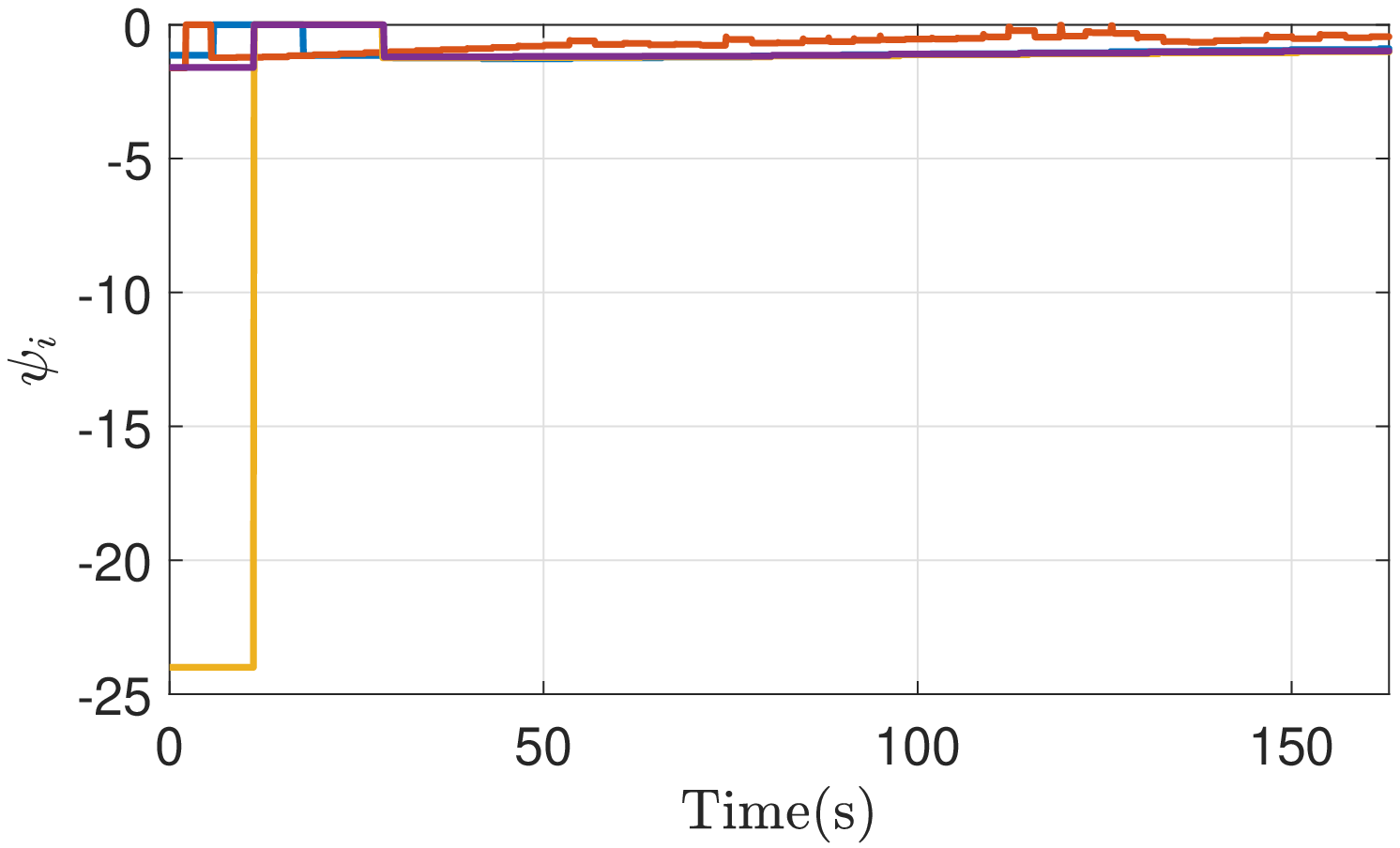}
\caption{Cost function of bargaining game theory implementation in a non-symmetric case.}
\label{fig:5-5a}
\end{minipage}
\quad
\begin{minipage}[b]{0.3\linewidth}
\includegraphics[width = \textwidth]{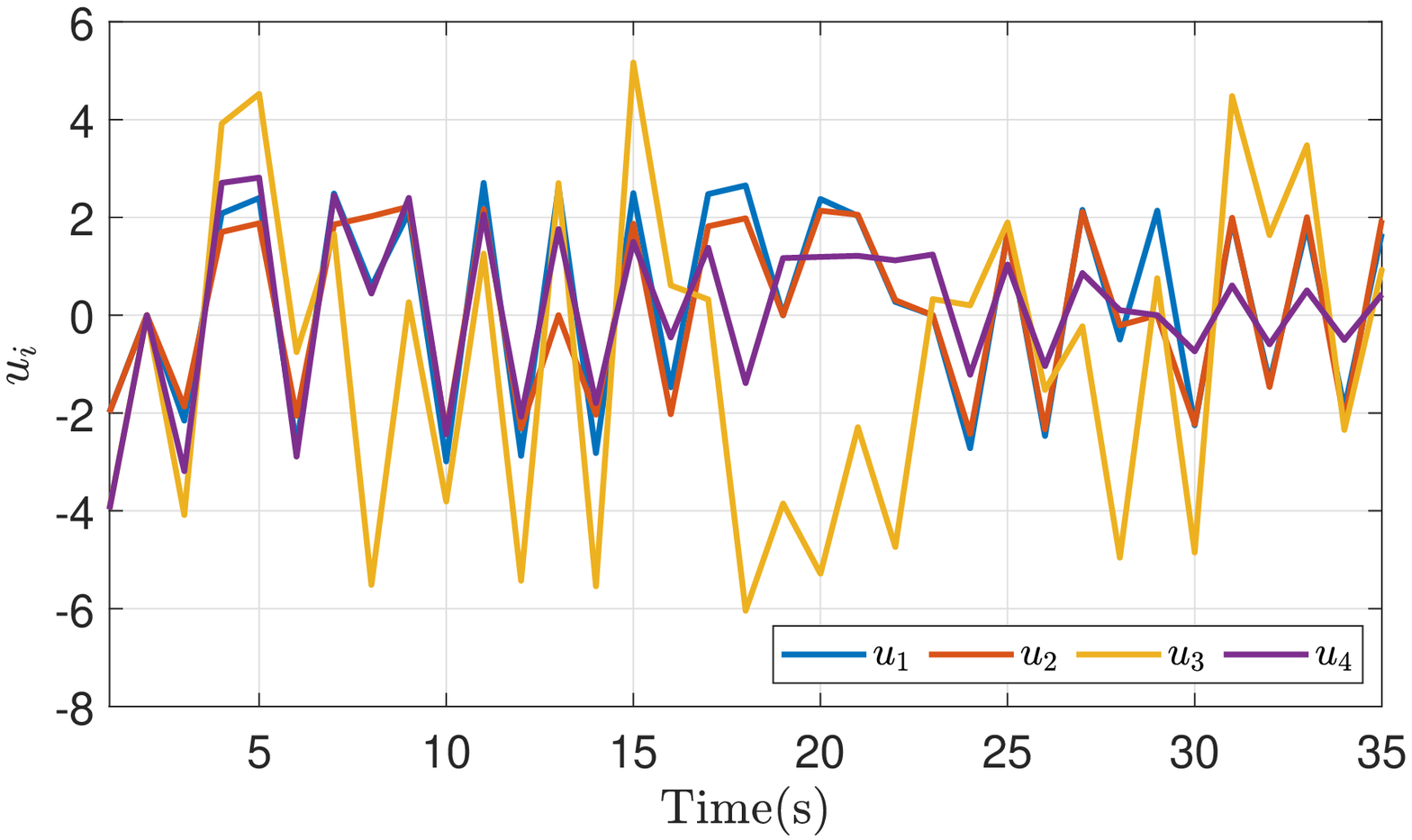}
\caption{Control action of bargaining game theory implementation in a non-symmetric case.}
\label{fig:5-5b}
\end{minipage}
\end{figure}

Figures \ref{fig:5-5}, \ref{fig:5-5a} and \ref{fig:5-5b} show the response of the point of disagreement, the cost function, and the control action in the non-symmetrical case of emulation, the response as in the previous cases, shows a fluctuation in their behavior, followed by a correct bargain in both cases.

\begin{figure}[!htb]
	\centering
	\includegraphics[width = 0.47 \textwidth]{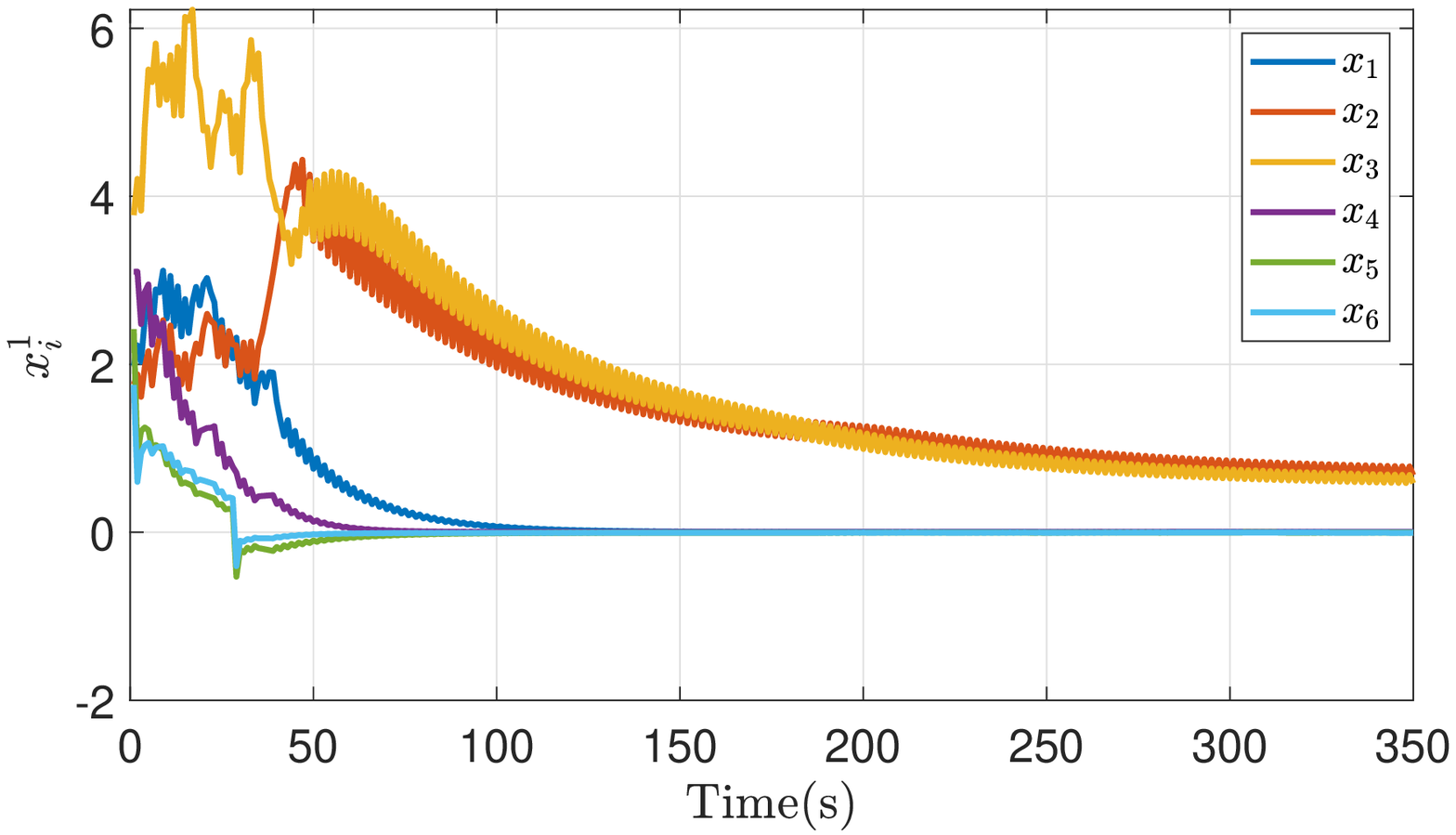}
	\caption{Output synchronization of bargaining game theory implementation in a non-bargaining case.}	
	\label{fig:5-6}
\end{figure}

In the same way, it is important to validate cases where a correct bargain is not achieved. Considering that, a scenario of multiple agents in the context of mechanical systems based on \cite{Arevalo2020adaptive}, presents a non-linearity in its systemic base that makes it difficult for the bargain algorithm, as is observed in Fig. \ref{fig:5-6}. It is validated that in those scenarios, when the value of the point of disagreement diverges, synchronization of all the systems is not completely achieved.

\section{Conclusion} \label{S6}

This paper solves the problem of the cooperative cruise in vehicles from a control perspective. Its logic is validated through predictive control in output synchronization of symmetric and non-symmetric agents with linear dynamics. The use of algorithms based on game theory allows conceptualizing the term cooperation and its relation with control through the Nash theory of bargaining, which is validated throughout the paper. 
A bargaining game methodology is used as a model predictive control strategy for game theory analysis. Its extension to dynamic cases is presented even in a discrete perspective for implementation. In this methodology, the analysis of the disagreement point is used dynamically for negotiation between agents, guaranteeing that this can be applied for symmetric and non-symmetrical games. The unique solution can be summarized as a convex optimization problem methodology.

At the simulation level, under the context of cooperative cruise control, symmetric and non-symmetric cases are defined, and it is possible to demonstrate the synchronization and the agreement reached through the bargaining algorithm. In the cases of non-symmetric games, the system tends to have a slower response and agreement derived from its characteristics. Similarly, in implementation, by emulating dynamic systems in hardware-in-the-loop, the dynamics of the agents are synchronized, achieving an agreement, highlighting that in non-symmetric games, the system presents fluctuations in its distributed response.

In future work, it is possible to strengthen this theory by including security parameters in cost functions and unstable models in an open loop, likewise, with the inclusion of heterogeneity parameters to have a better response in non-symmetric game implementation.


%
\section*{Declarations}
The development is partially financed by the project "RESEARCH PROGRAM IN EMERGING TECHNOLOGIES FOR SMART ELECTRIC MICROGRIDS WITH HIGH PENETRATION OF RENEWABLE ENERGIES" from Minciencias, Bogota, Colombia.

The authors declare that they have no conflict of interest.

The authors declare the transparency of the information and data obtained in the development of the article.

The authors confirm the availability of the codes without any inconvenience requested by email.

This work does not contain any studies with human or animal participants performed by any of the authors.

Informed consent was obtained from all individual participants included in the study.

Publication consent was obtained from all individual participants included in the study.

\bibliographystyle{spphys}       
\bibliography{main.bib}   

%
%

\end{document}